\documentclass[letterpaper]{article} 
\usepackage{aaai23}  
\usepackage{times}  
\usepackage{helvet}  
\usepackage{courier}  
\usepackage[hyphens]{url}  
\usepackage{graphicx} 
\usepackage{natbib}  
\usepackage{caption} 
\usepackage{algorithm}
\usepackage{algorithmic}

\usepackage{newfloat}
\usepackage{listings}
\DeclareCaptionStyle{ruled}{labelfont=normalfont,labelsep=colon,strut=off} 
\floatstyle{ruled}
\newfloat{listing}{tb}{lst}{}
\floatname{listing}{Listing}
\usepackage{xcolor}
\newcommand{\ie}{\textit{i.e., }}
\newcommand{\eg}{\textit{e.g., }}

\usepackage{amsmath}
\usepackage{amsfonts}
\usepackage{amsthm}
\usepackage{amssymb}
\usepackage{mathtools}
\usepackage{subcaption}
\newtheorem{theorem}{Theorem}

\newtheorem{lemma}{Lemma}
\newtheorem{definition}{Definition}

\title{Rethinking Safe Control in the Presence of Self-Seeking Humans}
\author {
    Zixuan Zhang\equalcontrib,
    Maitham AL-Sunni\equalcontrib,
    Haoming Jing\equalcontrib,
    Hirokazu Shirado,
    Yorie Nakahira
}
\affiliations {
    Carnegie Mellon University\\
    \{zzhang4, malsunni, haomingj\}@andrew.cmu.edu, \{shirado, yorie\}@cmu.edu
}

\begin{document}

\maketitle

\begin{abstract}
Safe control methods are often intended to behave safely even in worst-case human uncertainties. However, humans may exploit such safety-first systems, which results in greater risk for everyone. Despite their significance, no prior work has investigated and accounted for such factors in safe control. In this paper, we leverage an interaction-based payoff structure from game theory to model humans’ short-sighted, self-seeking behaviors and how humans change their strategies toward machines based on prior experience. We integrate such strategic human behaviors into a safe control architecture. As a result, our approach achieves better safety and performance trade-offs when compared to both deterministic worst-case safe control techniques and equilibrium-based stochastic methods. Our findings suggest an urgent need to fundamentally rethink the safe control framework used in human-technology interaction in pursuit of greater safety for all.
\end{abstract}

\section{Introduction}

This paper focuses on the safety-critical interactions of human agents and autonomous agents in the mixture of self-seeking and altruistic behaviors. Many safe control methods intend to behave safely in worst-case human uncertainties. The uncertainties can be great due to the complexity and difficulty of modeling human behaviors, which forces these methods to maintain a large safety margin and behave conservatively. 
However, when we assume humans are self-seeking actors, the standard safe methods might elicit the opposite effect. For example, when a human driver realizes that autonomous vehicles (AVs) always yield their right of way, the driver may cut in and change lanes aggressively toward AVs, which would pose greater risks for everyone. Safe control technology could be improved by incorporating how humans behave in response to autonomous systems.

\begin{figure}[h]
    \centering
    \includegraphics[width=\linewidth]{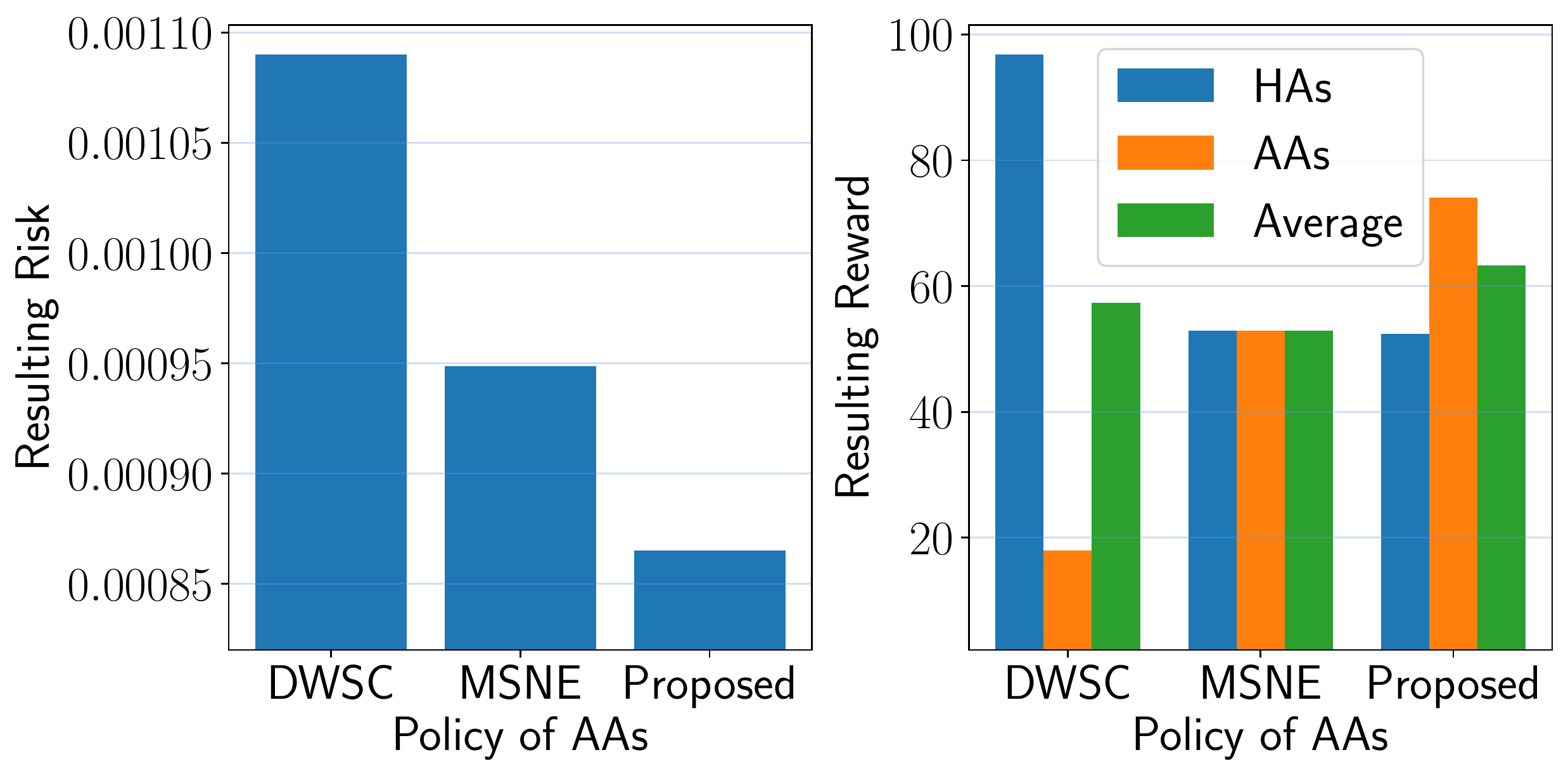} 
    \caption{Resulting risk (left) and reward (right) when Autonomous Agents use different safe control methods in a Type A interaction. Types of interactions are defined in Section~\ref{sec:reward_risk_categories}.}
    \label{fig:compare_risk_reward_type_A}
\end{figure}

In this work, we investigate the impact of strategic human behaviors in such interactions using an integration of control- and game-theoretic models. We use the principles of evolutionary game to characterize how rational humans change their strategy and behaviors over time and their impact on the performance and safety of the interaction.
We classified diverse scenarios into four qualitatively different types and studied when safety is intended for worst-case human uncertainties, denoted as deterministic worst-case safe control (DWSC), and when equilibrium-based stochastic strategies, denoted as mixed strategy Nash equilibrium (MSNE), are chosen. Interestingly, the deterministic safe control discourages collaborative human behaviors, resulting in more risky interactions (Lemma \ref{lem:dwsc_not_safe}, Figure \ref{fig:compare_risk_reward_type_A}). 

\begin{figure}[h]
    \centering
    \includegraphics[width=\linewidth]{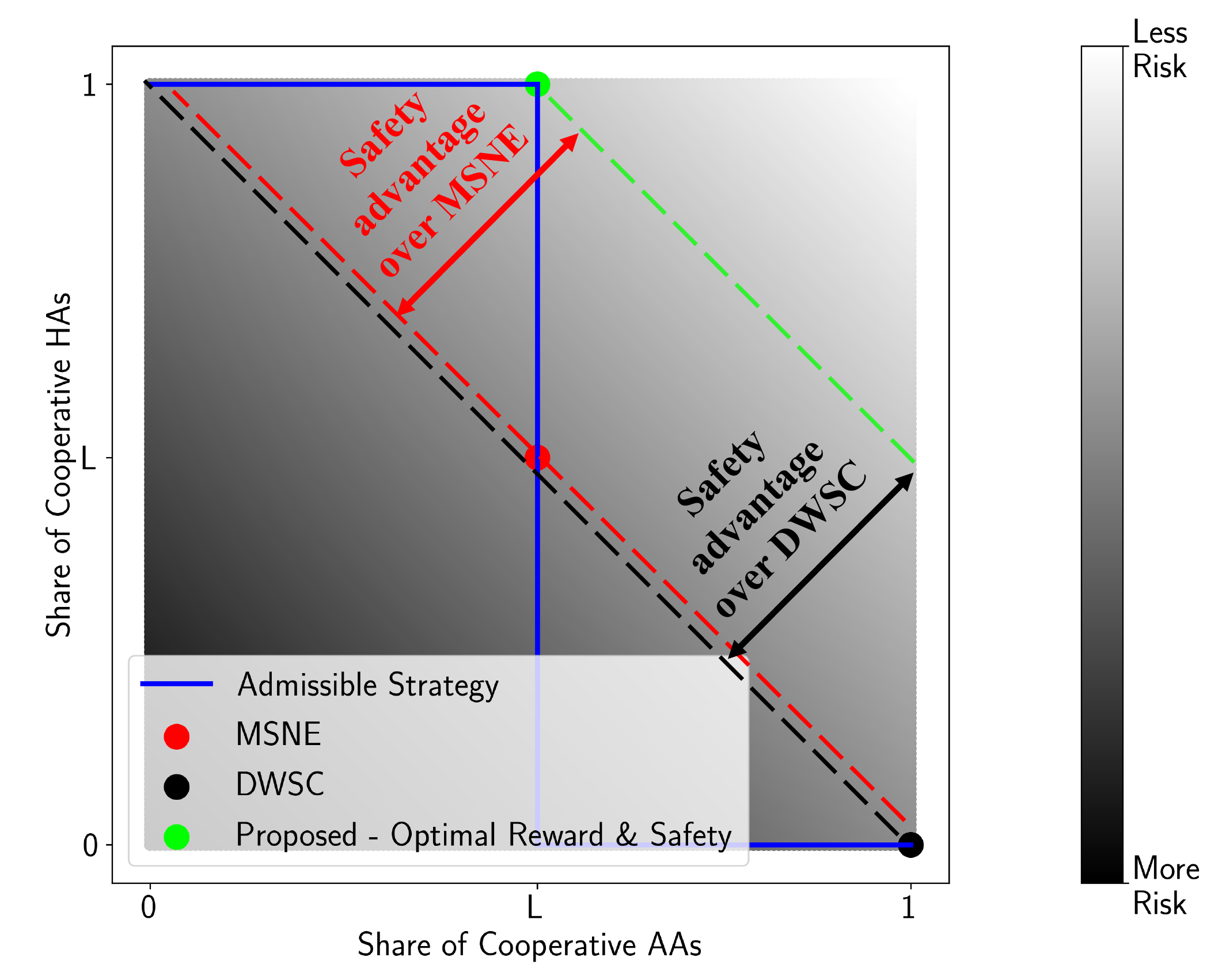}
    \caption{Risk map in a Type A interaction. The blue line shows the strategies achievable by the proposed method. The dashed lines are level sets for the risk. The proposed method results in less risk compared to MSNE and DWSC. Types of interactions are defined in Section~\ref{sec:reward_risk_categories}.}
    \label{fig:risk_gradient}
\end{figure}

Building on these insights, we then propose a method to encourage human to behave in an optimum way such that the safety of the overall system is maximized (Theorem \ref{thm:proposed_method_combined}, Figure \ref{fig:compare_risk_reward_type_A}, Figure \ref{fig:res_obj_1}). The proposed policy has better safety and performance trade-offs when compared to both deterministic worst-case safe control methods and equilibrium-based stochastic methods in the presence of strategic human behaviors (Figure \ref{fig:compare_risk_reward_type_A}, Figure \ref{fig:res_obj_2a}). The advantage of the proposed method in risk management compared to MSNE and DWSC for one type of interactions is shown in Figure \ref{fig:risk_gradient}.

\section{Related Work}
\subsubsection{Safe Control}

Many safe control methods exist for the design of autonomous systems that interact with humans. Some model human behaviors as uncertainties and noises and use stochastic safe control and multi-agent control~\cite{ahmadi2019safe,luo2020multi,lyu2021probabilistic,cheng2020safe}. Others use various human models~\cite{kulic2007affective,ding2011human,kelley2008understanding,ravichandar2018gaze,koppula2015anticipating} to design control policies in a variety of tasks: \eg robotic swarm control~\cite{atman2018motion,diaz2017human}, manipulation tasks~\cite{erhart2016model,peng2018collaborative} and autonomous vehicle control~\cite{cummings2011impact}. These methods are often designed to accommodate human behaviors and act with large safety margins with the intent to reduce tail (risk) events in the mixture of autonomous and human-driven cars. Since people use social information about others in coordinated movements~\cite{Faria2010-uq}, however, human drivers can develop risky behavior when they learn their counterpart is ``playing the coward." In short, the cooperative policies of autonomous systems can be simply exploited by non-cooperative people (\eg~the people who pursue self-interests)~\cite{Ishowo-Oloko2019-vn,Dawes1980-il,Shirado2020-hs} and, as a result, may lead to greater risks for the entire system. To avoid such unintended consequences and improve the safety of both humans and machines, we might need to account for social interactions between them~\cite{Chen2017-ns} in the safe control framework, which is the focus of this paper.

\subsubsection{Human-Machine Cooperation}

Game theory is one of the major theoretical frameworks to examine complex social interactions. Using the framework, researchers have studied how cooperation can emerge from rational actors~\cite{Axelrod1984-pg}. Cooperation is actually challenging because it creates a social dilemma (also known as the free-rider problem) ~\cite{Dawes1980-il}. A group does well if individuals cooperate, but each individual is tempted to defect~\cite{Olson1965-bc}. Even if one individual cooperates with others, the others could have an easy life by exploiting the first individual's benevolent effort~\cite{Nowak2006-bl}. To overcome such cooperation dilemmas, a large body of work has explored broader, institutional approaches, such as punishment~\cite{Fehr2002-ur}, group dynamics~\cite{Shirado2013-nl}, and the establishment of a central authority~\cite{Ostrom1990-wp}. 

The cooperation problem also occurs in mixed groups of humans and machines. For example, Shirado and Christakis have introduced preprogrammed autonomous agents (bots) into a network of people to examine which bot strategies can facilitate cooperation in human groups~\cite{Shirado2020-hs}. In the study, the bots that always cooperated with humans were simply exploited by them, and most people eventually chose defection with the cooperative bots. Ishowo-Oloko et al. shows that people do not cooperate, especially when they realize that they are interacting with autonomous systems ~\cite{Ishowo-Oloko2019-vn}. As theoretical and empirical evidence suggests the importance of accounting for self-seeking behaviors in cooperation, machines need to consider such human nature to facilitate cooperative human-machine systems~\cite{Paiva2018-iu, Dafoe2021-cw,Rahwan2019-rk}. This paper explores this implication in the safe control framework.

\section{Problem Statement}
\label{sec:ps}
\subsection{System Model}
We apply an interaction model of evolutionary game theory that two types of agents, Human Agents (HAs) and Autonomous Agents (AAs) interact with each other based on their payoffs for infinite periods~\cite{Hofbauer1998-jg, Nowak2006-bl}. Specifically, the interactions between HAs and AAs are modeled as follows. We assume the existence of infinitely many HAs and AAs, and focus on the interaction between a HA and an AA.  Their decision models are composed of interaction strategy $\pi=(\pi^h,\pi^a)$ and control policy $\phi = (\phi^h, \phi^a)$.  Throughout this paper, we use superscript $h$ to denote HAs and superscript $a$ to denote AAs. 

At each interaction of a HA and an AA, their intention $I$ is decided based on their strategies $\pi$ as follows:
\begin{align}
\label{eq:strategy}
    \pi^h &= \mathbb{P}(I^h=C)=1-\mathbb{P}(I^h=D) \\
    \pi^a &= \mathbb{P}(I^a=C)=1-\mathbb{P}(I^a=D),
\end{align}
The intentions can either be conservative (denoted as $C$ for cooperative) or aggressive (denoted as $D$ for defect), \ie $I^h, I^a \in \{C,D\}$. 

Given the intention, they use the control policy
\begin{align}
 &\mathbb{P}(u^h_{[k]}| I^h,x_{[k]}) \\
 &\mathbb{P}(u^a_{[k]}|I^a,x_{[k]}),
\end{align}
where $k \in \{0,1,\cdots,K\}$, to generate the control action $u = (u^h, u^a)$ based on the state of both agents. 
Here, we use subscript $[k]$ to denote the discrete time point $k\Delta t$, where $\Delta t$ is the sampling interval. Here, $ u = \{ u_{[k]} = [u_{[k]}^{h} ,u_{[k]}^{a}]^T , k \in \{0,1,\cdots,K\} \} $ and $x = \{ x_{[k]} = [x_{[k]}^{h} ,x_{[k]}^{a}]^T , k \in \{0,1,\cdots,K\} \}$ are the control action and state of the HA and the AA at time $k$, and $\{0,1,\cdots,K\}$ is the duration of interaction. We assume the control policy is identical among the population.

The system dynamics is characterized by the conditional transition probability $\mathbb{P}(x_{[k+1]}|x_{[k]},u_{[k]})$, $k \in \{0,1,\cdots,K\}$, which is identical among the population\footnote{The system dynamics is assumed to be uniform for both the HA and the AA.}. We will quantify the reward of each agent and the performance and safety of an interaction as follow. When the states and control actions end up being $x$ and $u$, the reward of the HA and the AA is given by $\rho^h(x,u)$ and $\rho^a(x,u)$. Let 
\begin{align}
    R^h&=\mathbb{E}[\rho^h(x,u)]\\
    R^a&=\mathbb{E}[\rho^a(x,u)].
\end{align}
denote the expected reward received by the HA and the AA. The expected performance of the interaction is given by
\begin{align}
    R=R^h+R^a.
\end{align}
In addition to the reward, a latent risk is also present with the interaction. 
The risk $W$ is quantified by the probability of the occurrence of some undesirable risk event, denoted as $\mathcal{U}$, \ie
\begin{align}
    W=\mathbb{E}[\mathbb{P}(\mathcal{U}|x,u)].
\end{align}
The risk we consider here is the types of risks that are not the major decision factors of humans, such as the long-term future risk, which are usually downplayed by human against the immediate reward~\cite{shirado2020collective}. The risks that are major decision factors of humans are incorporated into the reward. 

HAs and AAs use different sets of information about the outcome of past interactions to change their strategies based on the outcome. We assume that the strategy update is performed at a sufficient slower timescale than individual interactions, so the strategy update can use accurate statistics of the outcomes associated with the past and current strategy\footnote{Here, we assume that humans collect sufficient information (interaction samples) before they change their behaviors. For example, if a HA meets with an AA showing conservative behaviors, it will not exploit this behavior. However, if the HA meets with the AA for sufficiently many times that it can confirm the AA will always act conservatively, it will starts to exploit the conservativeness with aggressive behaviors.}. As a result, we use different notations for the interaction time and the strategy update time, \ie subscript $[k]$ for interaction time and subscript $t$ for strategy update time. Each agent will have the information about the expected reward they receive under certain intentions. The AAs will have the information of the total reward $R$ of the system as well. Unlike reward, only the AAs are able to calculate the latent risk $W$. This information asymmetry models the following two factors: the strategy of an AA can use the aggregate information of all other AAs; in contrast, HAs may not have a good estimate of the rare event probability such as crashes based on their experience, and HAs who get into accidents may exit from the population. 

We model human behaviors based on the widely accepted framework of myopic and bounded rationality in a distributed coordination, where HAs choose whether to cooperate based on the expectation of self-interests in the short term, \ie the individual reward $R^h$. In this setting, humans are well modeled as ``conditional cooperators'' theoretically and empirically~\cite{hilbe2013evolution,nowak2005evolution}. In this setting, there exists a few common ways from existing literature that models the strategy update of HAs in evolutionary game theory~\cite{Hofbauer1998-jg, Nowak2006-bl}. These models estimate possible outcomes from counterpart moves in different ways. To account for a wide range of possibilities, we adopt 3 of such models and consider an update rule consisting a mixture of these models. The human strategy update rule for each of the models are defined below.
\begin{itemize}
    \item Replicator Dynamics~\cite{replicator}.
    \begin{align} \label{eq:rep_dyn}
        \dot{\pi}^h_t=&\pi^h_t(\mathbb{E}[R^h|\pi^h_t=1,\pi^a_t]-\mathbb{E}[R^h|\pi^h_t,\pi^a_t])\nonumber\\
        :=&f_r(\pi^h_t,\pi^a_t).
    \end{align}
    \item Brown-Nash-von Neumann Dynamics~\cite{bnn}.
    \begin{align} \label{eq:bnn_dyn}
        \dot{\pi}^h_t=&[\mathbb{E}[R^h|\pi^h_t=1,\pi^a_t]-\mathbb{E}[R^h|\pi^h_t,\pi^a_t]]_+\nonumber\\
        &-\pi^h_t([\mathbb{E}[R^h|\pi^h_t=1,\pi^a_t]-\mathbb{E}[R^h|\pi^h_t,\pi^a_t]]_+\nonumber\\
        &-[\mathbb{E}[R^h|\pi^h_t=0,\pi^a_t]-\mathbb{E}[R^h|\pi^h_t,\pi^a_t]]_+)\nonumber\\
        :=&f_b(\pi^h_t,\pi^a_t).
    \end{align}
    Here, $[q]_+=\max(0,q)$.
    \item Smith Dynamics~\cite{smith}.
    \begin{align} \label{eq:smith_dyn}
        \dot{\pi}^h_t=&(1-\pi^h_t)[\mathbb{E}[R^h|\pi^h_t=1,\pi^a_t]-\mathbb{E}[R^h|\pi^h_t=0,\pi^a_t]]_+\nonumber\\
        &-\pi^h_t[\mathbb{E}[R^h|\pi^h_t=0,\pi^a_t]-\mathbb{E}[R^h|\pi^h_t=1,\pi^a_t]]_+\nonumber\\
        :=&f_s(\pi^h_t,\pi^a_t).
    \end{align}
    Here, $[q]_+=\max(0,q)$.
\end{itemize}
The mixed dynamics update rule is given by
\begin{align}
\label{eq:mixed_update_rule}
    \dot{\pi}^h_t=&w_rf_r(\pi^h_t,\pi^a_t)+w_bf_b(\pi^h_t,\pi^a_t)+w_sf_s(\pi^h_t,\pi^a_t)\nonumber\\
    :=&f_m(\pi^h_t,\pi^a_t),
\end{align}
where $w_r$, $w_b$ and $w_s$ are the weights for Replicator Dynamics, Brown-Nash-von Neumann Dynamics and Smith Dynamics, respectively, and 
\begin{align}
    w_r+w_b+w_s=1.
\end{align}
Here, we make a reasonable assumption that $\pi^h_0\neq 0$ and $\pi^h_0\neq 1$. In other word, we assume that at the start, HAs are not entirely cooperative or entirely defect.


Our objective is to optimize the performance of the interaction while controlling the risk to be within a tolerable range. Toward this goal, we will design the strategy update rules of the AAs, which in turn influence the strategy of HAs, for optimizing the outcome of the AA-HA interactions, which depends on the strategies of both AAs and HAs. This objective is formally stated below. 
\begin{align}
\label{eq:design_objective}
    \pi^\star=\underset{\pi\in\mathcal{A}}{\arg\max}\ &\mathbb{E}[R|\pi]\\
    \textrm{subject to }&\mathbb{E}[W|\pi]\leq \epsilon.\nonumber
\end{align}
Here, $\epsilon$ is the tolerable risk, we assume it is chosen such that \eqref{eq:design_objective} is feasible, and $\mathcal{A}$ is the admissible strategy set, which is defined below.
\begin{definition}[Admissible Strategy Set]
\label{def:admissible_strategy}
The admissible strategy set is the set of all strategies $\pi$ that make HAs' strategy remain static, \ie $\dot{\pi}^h=0$.
\end{definition}
Here, the admissible strategy is not equivalent to the equilibrium. In fact, an equilibrium is an admissible strategy, but not every admissible strategy is an equilibrium. An equilibrium is a point where both $\pi^h$ and $\pi^a$ do not move. On the other hand, an admissible strategy only requires $\pi^h$ to stop evolving, since $\pi^a$ is something we have control on, it can be either static or dynamic, and it is decided by any control policy. As stated in \eqref{eq:design_objective}, our design objective is to achieve an optimum policy within the set of admissible strategy. We consider this limitation because it can be difficult to design a control policy when HAs are changing strategies. In fact, most applied control policies are designed based on a training distribution (i.e., a certain HAs environment).

\subsection{Reward and Risk Categorization}
\label{sec:reward_risk_categories}
We use a notation system such that $R_{XY}$ denotes the expected reward when an agent chooses strategy $X$ and its confronting agent chooses strategy $Y$. For simplicity, we consider a symmetric reward table. Likewise, for risk, we consider the same notation system. However, unlike reward, here we have one risk quantity for each interaction. Also, we require a few reasonable assumptions:
\begin{gather}
    W_{CD}=W_{DC} \\
    R_{DD}<R_{CD}\label{eq:assumption2}\\
    R_{CC}<R_{DC}.\label{eq:assumption3}
\end{gather}

The simplified reward and risk table is given in Table~\ref{tab:general_reward}.
\begin{table}[H]
    \centering
    \begin{tabular}{c|c|c}
         & $C$ & $D$ \\\hline
        $C$ & ($R_{CC}$, $R_{CC}$);$W_{CC}$ & ($R_{CD}$, $R_{DC}$);$W_{CD}$ \\
        $D$ & ($R_{DC}$, $R_{CD}$);$W_{CD}$ & ($R_{DD}$, $R_{DD}$);$W_{DD}$ 
    \end{tabular}
    \caption{Reward and risk table.} 
    \label{tab:general_reward}
\end{table}
We typically observe scenarios whose reward and risks are ordered as follows. 
\begin{align}
    \textrm{Reward case 1: }&2R_{CC}>R_{CD}+R_{DC}>2R_{DD} \label{eq:reward_case1}\\
    \textrm{Reward case 2: }&R_{CD}+R_{DC}>2R_{CC}>2R_{DD} \label{eq:reward_case2}\\
    \textrm{Risk case 1: }&W_{CC}<W_{CD}<W_{DD}\\
    \textrm{Risk case 2: }&W_{CD}<W_{CC}<W_{DD}.
\end{align}
The interactions of the above case is qualitatively different. To better characterize such differences, we classify the interactions as below, along with example scenarios in the autonomous driving settings.
\begin{itemize}
    \item Type A: reward case 1 and risk case 1.
    \item Type B: reward case 1 and risk case 2. 
    \item Type C: reward case 2 and risk case 1.
    \item Type D: reward case 2 and risk case 2.
\end{itemize}

Intuitively, when the reward and risk belong to the same case, there exist strategies that simultaneously maximizes the reward and minimizes the risk. In this scenario, the optimizer of \eqref{eq:design_objective} is such strategy. When the reward and risk belong to different cases, the two objectives compete. In this scenario, the optimizer of \eqref{eq:design_objective} for varying risk tolerance $\epsilon$ characterizes a set of pareto-optimal strategies.

The above model account for typical characteristics in human-machine interactions and has the following distinct factors from conventional game-theoretic models. First, there is a safety (or a latent risk) factor in addition to the rewards. Second, the available information is asymmetric: HAs can only estimate the expected rewards, while AAs can estimate both expected rewards and risks. Third, HAs is self-seeking while AAs is designed to optimize the safety and reward of the whole system. In the next section, we will understand the influence of short-sighted self-seeking humans, and study how to account for such propensity for improving the safety and efficacy in collective movements.

\section{Case Studies} \label{case_studies}
\subsection{Autonomous Driving Simulation} \label{sec:autonomous_driving_simulation}
\begin{figure}
    \centering
    \includegraphics[width=\linewidth]{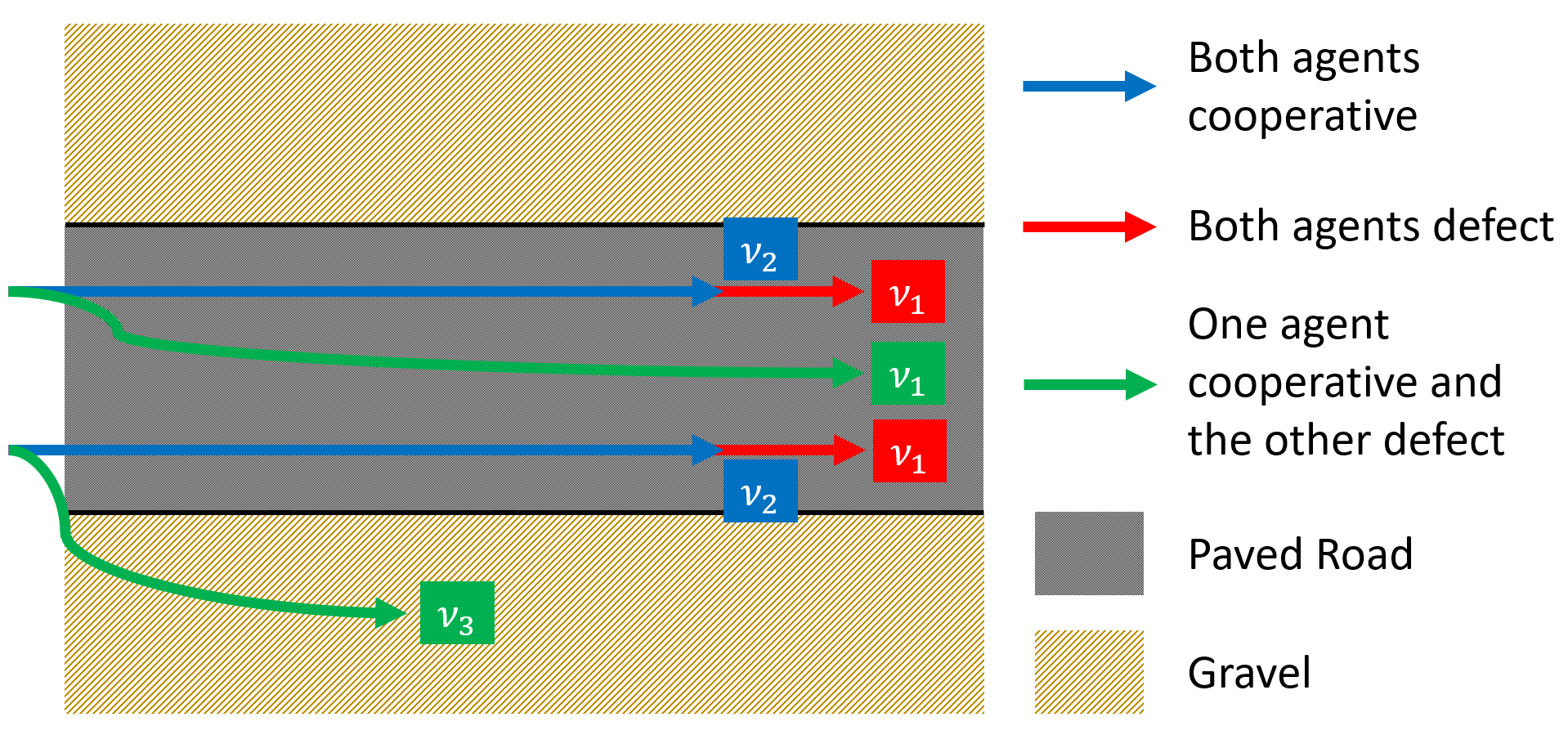}
    \caption{Simulated driving scenario.}
    \label{fig:reward_sim}
\end{figure}

We use some typical autonomous driving settings to study the types of interactions. Typical scenarios for each type in the autonomous driving setting are as follows:
\begin{itemize}
    \item Type A: At a stop-sign controlled intersection, where the aggressive behavior of one vehicle leads to less total reward (causing a havoc in passing order that takes time to be resolved) and more risk (crashing into vehicles passing normally).
    \item Type B: Driving on a narrow road that constantly poses hazards to the vehicles on it (\eg falling rock). The aggressive behavior of one vehicle helps reduce the risk by deciding the passing order in a time shorter than the time needed for 2 cooperative vehicles to negotiate the order. 
    \item Type C: In a lane-changing scenario, the aggressive behavior of one vehicle creates more risk since it is more likely to crash by not yielding. On the other hand, it gives better total reward by eliminating its yield time.
    \item Type D: In a high speed lane-changing scenario. Different from the previous scenario, yielding and reducing speed when many vehicles are driving at high speed may leads to more likelihood for crashing.
\end{itemize}
We use a narrow road driving example illustrated in Figure \ref{fig:reward_sim} to simulate the realistic reward and risk values that represents all 4 types. The example considers two vehicles, one is driven by a human and the other is an AV, on a narrow road with gravel outside of the paved road. When driving on the gravel, the vehicles will have a crash probability. The AV has the option to act cooperatively by decelerating and adopting a DWSC approach. This approach keeps a safety distance whose length depends on the speed of the other vehicle. It also has the option to defect by ignoring the safety distance. If both vehicles choose to cooperate, they will both decelerate to a speed $\nu_2$ such that they can both drive on paved road without violating the safety distance. If the human driver choose to defect, continuing driving with speed $\nu_1$ without decelerating, and the AV adopts DWSC, the AV will be forced onto gravel due to safety distance constraints and only drive at a much more reduced speed $\nu_3$. If both vehicles choose to defect, they will drive the paved road with their original speed $\nu_1$. The detailed specifications for the simulation is provided in the Appendix. Table \ref{tab:all_params} shows the generated rewards and risks for all cases.

\begin{table}[h]
  \begin{center}
    \begin{tabular}{c|c|c|c|c}
      & Type A & Type B & Type C & Type D \\
      \hline
      $R_{CC}$ & 65.51 & 53.53 & 60.29 & 56.13 \\\hline
      $R_{CD}$ & 17.93 & -0.05 & 40.79 & 49.87 \\\hline
      $R_{DC}$ & 96.8 & 68.7 & 95.28 & 88.24\\\hline
      $R_{DD}$ & -69.23 & -264.59 & 40.31 & 43.49 \\\hline
      $W_{CC}$ & 0.00078 & 0.00147 & 0.00058 & 0.00057 \\\hline
      $W_{CD}$ & 0.00109 & 0.00134 & 0.00073 & 0.00044 \\\hline
      $W_{DD}$ & 0.00147 & 0.00172 & 0.0015 & 0.00077
    \end{tabular}
    \caption{Reward and risk values for different types.}
    \label{tab:all_params}
  \end{center}
\end{table}

\subsection{Existing Methods}
In DWSC, AAs are designed to be always collaborative (\ie they adopt the time-invariant policy $\pi^a = 1$). The intention is to make AAs always ready for worst-case scenarios. However, such designs overlook the fact that in the presence of self-seeking HAs, the risk of the interaction might be higher than other not very conservative control policies. 

To investigate the safety of DWSC, we need to see the convergence of HAs' policy under DWSC.
\begin{lemma}
\label{lem:dwsc_not_safe}
Assuming \eqref{eq:assumption2} and \eqref{eq:assumption3}. If HAs are following the mixed dynamics \eqref{eq:mixed_update_rule}, an AA that will always choose $\pi^a = 1$  will result in $\pi^h_t \xrightarrow[]{t\rightarrow\infty} 0$.
\end{lemma}
The proof is provided in the Appendix.

Based on Lemma \ref{lem:dwsc_not_safe}, we can calculate the expected risk as
\begin{align}
    \label{eq:risk_calc}
    \mathbb{E}[W|\pi_t]&=W_{CC}\pi^h_t\pi^a_t+W_{CD}\pi^h_t(1-\pi^a_t)\\ &+W_{CD}(1-\pi^h_t)\pi^a_t+W_{DD}(1-\pi^h_t)(1-\pi^a_t). \nonumber
\end{align}
As $t \xrightarrow[]{} \infty$,  we have $\pi^a_t = 1$ and $\pi^h_t = 0$. Hence,
\begin{align}
    \mathbb{E}[W|\pi_t]= W_{CD}.
\end{align}
Given the above, we can see that DWSC will always give $\mathbb{E}[W|\pi_t]= W_{CD}$. However, $W_{CD}$ is not necessarily the minimum among \{$W_{CC}$, $W_{CD}$, $W_{DD}$\}. This also can be seen in the cases simulated in subsection \ref{sec:autonomous_driving_simulation}. This suggests that DWSC will not provide the safest behaviors in Type A and Type C interactions in the presence of strategic HAs' behaviors. In these types, DWSC will always start at a certain level of safety (given the underlying HAs cooperation distribution); however, when HAs start to exploit AAs cooperation, risk level goes higher. As shown in Figure \ref{fig:res_obj_1}, at the beginning, DWSC starts at a certain level of safety but risk increases as HAs cooperation decreases with time (HAs distribution changes toward a non-favorable manner w.r.t. safety). With this, we see how DWSC achieves short-sighted safety. Accordingly, people should be careful about that.


Alternatively, MSNE methods are designed to drive both AAs' and HAs' policies to the Mixed Strategy Nash Equilibrium (MSNE), which is defined below.
\begin{definition}[Mixed Strategy Nash Equilibrium]
In an interaction between AAs and HAs, although we have both rewards and risks, we define the mixed strategy Nash equilibrium (MSNE) only based on rewards as
\begin{equation} \label{eq:msne}
L = \frac{R_{DD}-R_{CD}}{R_{CC}+R_{DD}-R_{DC}-R_{CD}}.
\end{equation}
\end{definition}
From \eqref{eq:assumption2} and \eqref{eq:assumption3}, we have
\begin{align}
\label{eq:assumption1}
    L\in(0,1).
\end{align}
In MSNE methods, risks and rewards are restricted to those that correspond to the equilibrium $\pi^a = \pi^h  = L$. These methods cannot achieve pareto optimality between performance and safety because it is restricted to the equilibrium strategy. 
Risk resulting from AAs adopting MSNE methods can be calculated using \eqref{eq:risk_calc} as
\begin{align}
    \mathbb{E}[W|\pi_t]&= L^2 (W_{CC} + W_{DD} - 2 W_{CD}) \\
    &+ 2L (W_{CD} - W_{DD}) + W_{DD}. \nonumber
\end{align}
Hence, no safety guarantee in MSNE methods, as its resulting risks mainly depend on different environmental variables. Figure \ref{fig:res_obj_1} shows safety and reward levels for MSNE methods compared with other methods in a Type A interaction.

\section{Proposed Algorithm}
\label{sec:proposed_algorithm}
In this section, we present our results and proposed algorithm with their theoretical guarantees.

Let
\begin{align}
    \mathcal{A}_0&=\{(\pi^h,\pi^a):\pi^h=0,\pi^a\in(L,1]\}\label{eq:start_admissible}\\
    \mathcal{A}_1&=\{(\pi^h,\pi^a):\pi^h=1,\pi^a\in[0,L)\}\\
    \mathcal{A}_d&=\{(\pi^h,\pi^a):\pi^h\in[0,1],\pi^a=L\}.
\end{align}
The set of admissible strategy $\mathcal{A}$ is given by
\begin{align}
\label{eq:end_admissible}
    \mathcal{A}=\mathcal{A}_0\cup\mathcal{A}_1\cup\mathcal{A}_d.
\end{align}
In Lemma 2, we show that the set $\mathcal{A}$ is indeed the admissible strategy as defined in Definition \ref{def:admissible_strategy}.
\begin{lemma}
\label{lem:stop}
Consider a system where AAs are adopting a policy $\pi^a_t$, HAs are following the mixed dynamics \eqref{eq:mixed_update_rule}. Assuming \eqref{eq:assumption2} and \eqref{eq:assumption3}, then, 
\begin{align}
\label{eq:threshold_conv}
    \pi_t:=(\pi^h_t,\pi^a_t)\in\mathcal{A}\Leftrightarrow\dot{\pi}^h_t=0.
\end{align}
\end{lemma}
The proof is provided in the Appendix.

Next, we present our proposed algorithm. Let
\begin{align}
    \mathcal{B}_0&=\left[\frac{\epsilon-W_{DD}}{W_{CD}-W_{DD}},\infty\right)\label{eq:begin_proposed_policy}\\
    \mathcal{F}_0&=\{(\pi^h,\pi^a):\pi^h=0,\pi^a\in(L,1]\cap\mathcal{B}_0\}\\
    \mathcal{B}_1&=\begin{cases}
        \left(-\infty,\frac{\epsilon-W_{CD}}{W_{CC}-W_{CD}}\right], & W_{CC}>W_{CD} \\ 
        \left[\frac{\epsilon-W_{CD}}{W_{CC}-W_{CD}},\infty\right), & W_{CC}<W_{CD}
    \end{cases}\\
    \mathcal{F}_1&=\{(\pi^h,\pi^a):\pi^h=1,\pi^a\in[0,L)\cap\mathcal{B}_1\}\\
    \mathcal{B}_d&=\begin{cases}
        \left(-\infty,\frac{\epsilon-W_{CD}L-W_{DD}(1-L)}{W_{CC}L+W_{CD}-2W_{CD}L-W_{DD}(1-L)}\right], \\ 
        \ \ \ \ W_{CC}L+W_{CD}-2W_{CD}L-W_{DD}(1-L)>0 \\ 
        \left[\frac{\epsilon-W_{CD}L-W_{DD}(1-L)}{W_{CC}L+W_{CD}-2W_{CD}L-W_{DD}(1-L)},\infty\right), \\
        \ \ \ \ W_{CC}L+W_{CD}-2W_{CD}L-W_{DD}(1-L)<0
    \end{cases}\\
    \mathcal{F}_d&=\{(\pi^h,\pi^a):\pi^h\in[0,1]\cap\mathcal{B}_d,\pi^a=L\}.
\end{align}
We define the feasible set of strategies under the constraints of \eqref{eq:design_objective}:
\begin{align}
\label{eq:end_feasible}
    \mathcal{F}= \mathcal{F}_0\cup \mathcal{F}_1\cup \mathcal{F}_d.
\end{align}
Then, the optimal strategy is given by
\begin{align}
    \pi^\star:=&(\pi^{h \star},\pi^{a \star})=\underset{(\pi^h,\pi^a)\in\mathcal{F}}{\arg\max}\mathbb{E}[R|\pi^h,\pi^a].\label{eq:pi_pound}
\end{align}
The proposed policy is given by
\begin{align}
\label{eq:end_proposed_policy}
    \pi^a_t=\begin{cases}
        L-G(\pi^{h \star}-\pi^h_t), & \pi^{\star}\in\mathcal{F}_d\\
        \pi^{a \star}, & \textrm{otherwise}
    \end{cases}:=f^a(\pi^h_t).
\end{align}
Here, $G \in \mathbb R$ is a strictly positive constant.
\begin{theorem} \label{thm:proposed_method_combined}
Consider a system where AAs are adopting the policy defined in \eqref{eq:begin_proposed_policy} to \eqref{eq:end_proposed_policy}, and HAs are following the mixed dynamics \eqref{eq:mixed_update_rule}. Assuming \eqref{eq:assumption2} and \eqref{eq:assumption3}, then, $(\pi^h_t,\pi^a_t) \xrightarrow[]{t\rightarrow\infty} (\pi^{h \star},\pi^{a \star})$, which is the solution to \eqref{eq:design_objective}.
\end{theorem}
The proof is provided in the Appendix.

The proposed algorithm is given in Algorithm \ref{alg:algorithm}.

\begin{algorithm}[tb]
\caption{Proposed algorithm.}
\label{alg:algorithm}
\textbf{Input}: Rewards and risks (Table \ref{tab:general_reward}), the tolerable risk $\epsilon$, the constant $G$.
\begin{algorithmic}[1] 
\STATE Compute $\mathcal{A}$ using \eqref{eq:start_admissible} to \eqref{eq:end_admissible}.
\STATE Compute $\mathcal{F}$ using \eqref{eq:begin_proposed_policy} to \eqref{eq:end_feasible}.
\STATE Compute $\pi^\star$ using \eqref{eq:pi_pound}.
\WHILE{$t>0$}
\STATE Observe HA strategy $\pi^h_t$.
\STATE $\pi^a_t\leftarrow f^a(\pi^h_t)$.
\ENDWHILE
\end{algorithmic}
\end{algorithm}

\section{Numerical Simulations}
\label{sec:simulation}

\begin{figure*}[h]
    \centering
    \includegraphics[width=\linewidth]{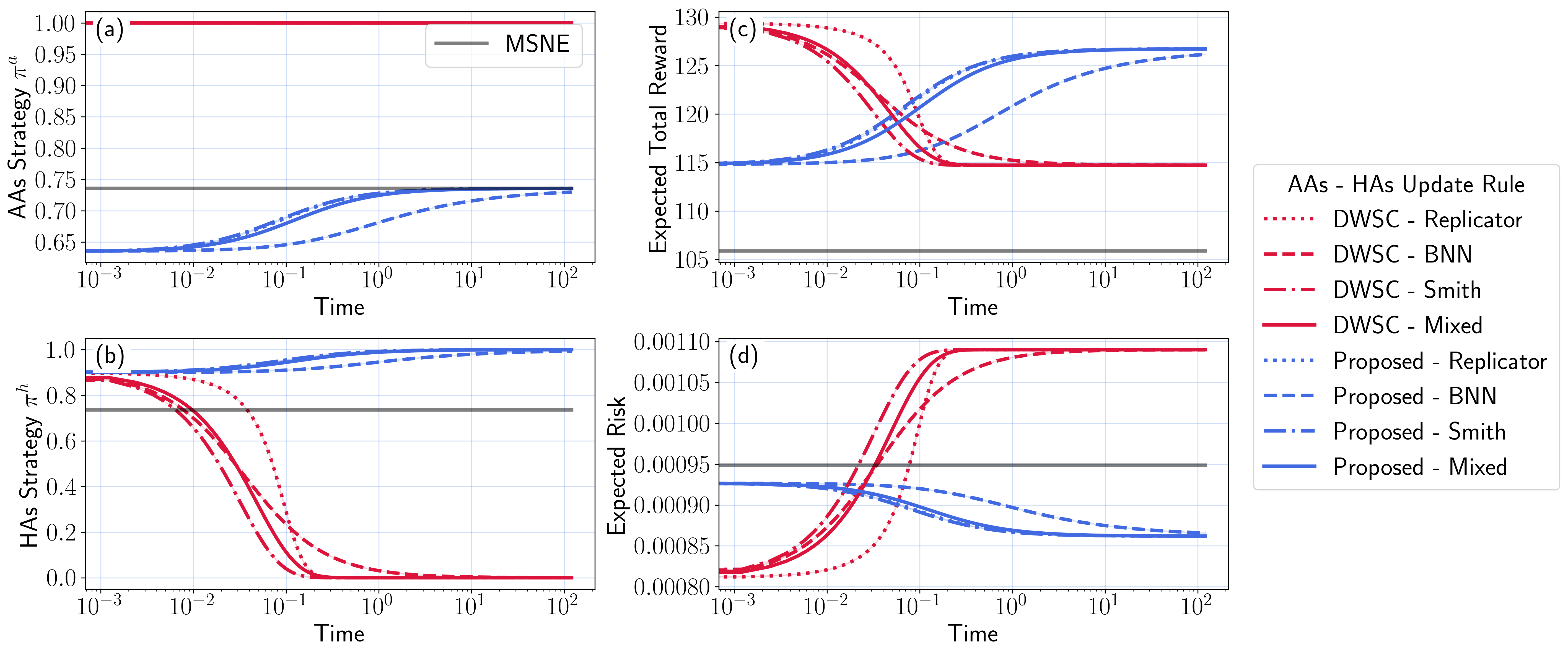}
    \caption{Results in a Type A interaction with different AAs-HAs update rules ($\pi^h_0=0.9$, $\epsilon=9e-4$, $G=1$ and $w_r=w_b=w_s=\frac{1}{3}$). In plot (b), HAs' strategies converge to $\pi^h=0$ if AAs takes DWSC. Meanwhile, the expected total reward in plot (c) decreases and the expected risk in plot (d) increases as $\pi^h_t \rightarrow 0$. Even though the initial rewards and risks of DWSC are quite desirable, it cannot remain static and it degrades quickly.
    In this case, the proposed method achieves the highest reward and lowest risk at the same time, while both the reward and risk that MSNE and DWSC achieve are sub-optimal.
    }
    \label{fig:res_obj_1}
\end{figure*}
\begin{figure*}[h]
    \centering
    \includegraphics[width=\linewidth]{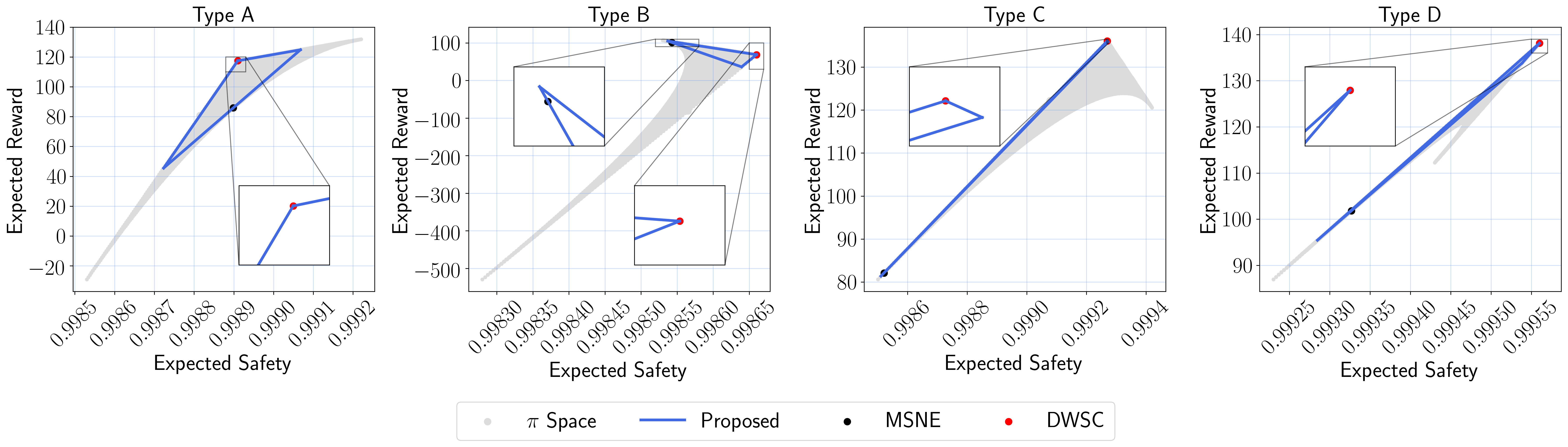}
    \caption{
    Illustration of achievable rewards and safety of different methods. The gray area labeled as `$\pi$ Space' denotes all possible interaction strategies. Note that the outcome of DWSC and MSNE methods are always achievable by the proposed method.}
    \label{fig:res_obj_2a}
\end{figure*}
\begin{figure}[h!]
    \centering
    \includegraphics[width=\linewidth]{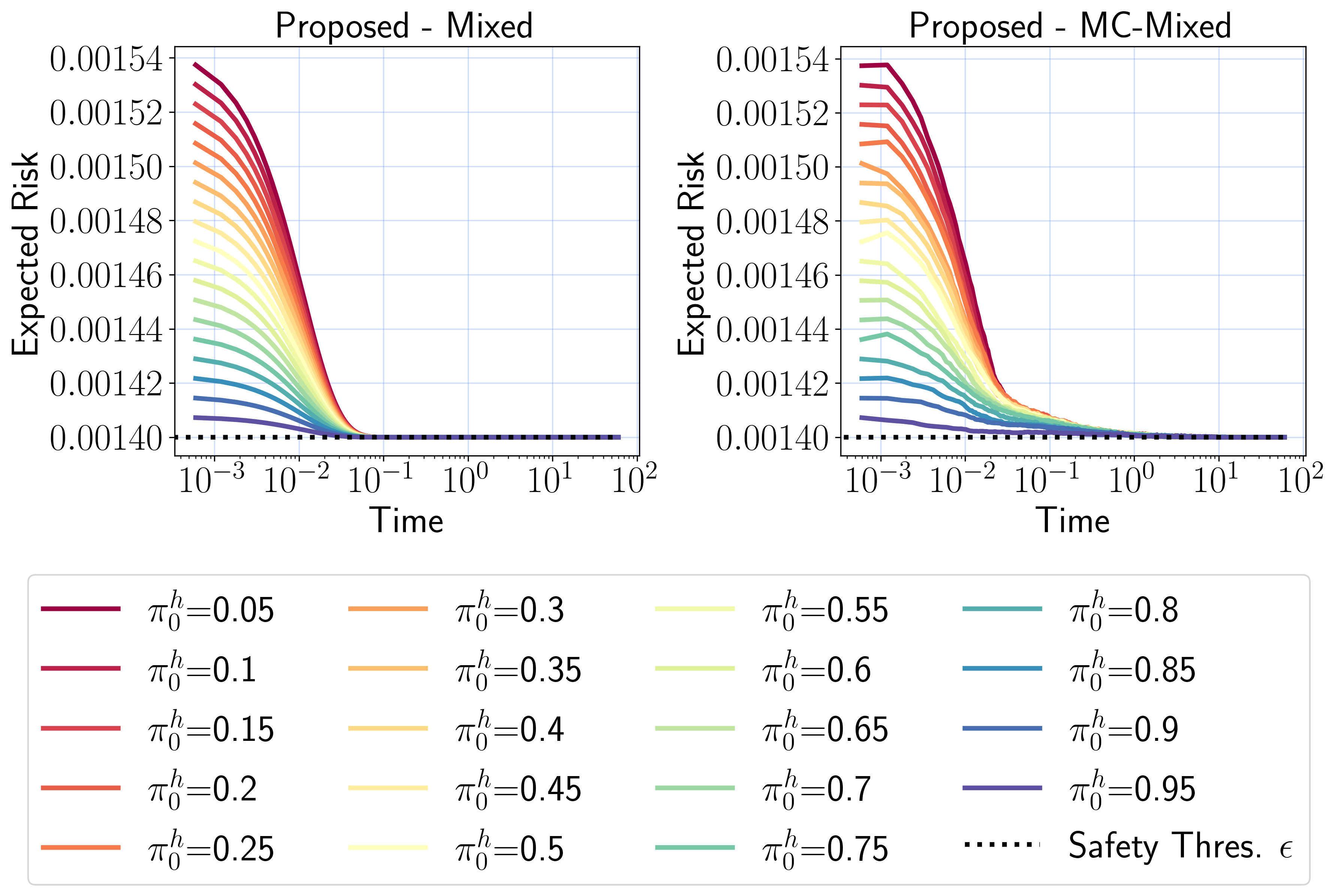}
    \caption{Robustness of the proposed methods controlling risks within a tolerable range against mixed HAs' update rules (\textbf{Left}: ODE, \textbf{Right}: Monte Carlo) given different initial conditions in a Type B interaction ($\epsilon=1.4e-3$, $w_r=w_b=w_s=\frac{1}{3}$ and $G=1$). 
    }
    \label{fig:res_obj_2b}
\end{figure}

The experiments are divided into two parts. First, we study the existing methods from DWSC to MSNE methods and aims to show that DWSC is not the safest and can result in riskier situations. Then we demonstrates the benefit of the proposed method that it can control risks within a tolerable range in the long time scale and achieve a better trade-off between safety and reward than both DWSC and MSNE. 

\subsection{Settings}
We performed numerical simulations in Python with both ODE-based update rules and Monte Carlo based update rules. 
We simulated the evolution of autonomous-human interactions with 4 types of interaction scenarios
introduced in Section \ref{sec:reward_risk_categories} 
with reward and risk table obtained in Section 
\ref{sec:autonomous_driving_simulation}. For HA update rules, we adopt both ODE based and Monte Carlo based versions of 4 dynamics, including Replicator Dynamics, Brown-Nash-von Neumann dynamics, Smith Dynamics, and a mixture of them.
For ODE based update rules, we update the states based on dynamics of the form \eqref{eq:rep_dyn} to \eqref{eq:mixed_update_rule} by the Runge–Kutta method \cite{runge1895numerische} over $1\times10^5$ time steps. 
In Monte Carlo simulations, we consider an infinite population of HAs and an infinite population of AAs interacting with each other. We randomly pick $N=1\times10^3$ pairs and update their intentions over $1\times10^5$ time steps. The proportion of cooperators in each sampled population is interpreted as the strategy $\pi$ for that population. Each individual updates its intention as follows,
\begin{itemize}
    \item \text{Replicator Dynamics (RD):} Each HA $i$ will randomly pick another HA $j$ with probability $\frac{1}{N-1}$, and change $I^h_i$ to $I^h_j$ with a probability proportional to the excess part of $j$'s reward over its own.
    \item \text{Brown-Nash-von Neumann Dynamics (BNN):} Each HA
    with intention $C$ will switch to $D$ with a probability proportional to \
    the excess part of $D$'s expected reward over HAs' average reward
    , and vice versa.
    \item \text{Smith Dynamics (SD):} Each HA with intention $C$ will switch to $D$ with a probability proportional to the excess part of $D$'s expected reward over $C$'s expected reward, and vice versa.
    \item \text{Mixed Dynamics:} It consists of three HA subpopulations that take RD, BNN, and SD respectively. Each individual will change their intention following the update rule of their subpopulation but considering the whole population when making updates rather than their own subpopulation.
\end{itemize}
For update rules of AAs, we compared the proposed methods with 
DWSC and equilibrium-based approaches. As discussed before, AA with DWSC will always be conservative for the worst-case, which means $\pi^a=1$ for all time. For equilibrium-based approaches, since they aims to achieve MSNE, we simulate the state of MSNE for reference.  

\subsection{Results}
We first compared the evolutionary trajectories of different AAs' update rules against different HAs' update rules. As Figure \ref{fig:res_obj_1} shows, taking DWSC and always being conservative results in relatively safe interactions at the beginning. Then, HAs' strategies converge to $\pi^h=0$ as time evolves, which corresponds to Lemma \ref{lem:dwsc_not_safe}. However, as $\pi^h\rightarrow 0$, the expected total reward decreases and the expected risk increases, and it eventually ends up in riskier interactions. That demonstrates the incapability of DWSC to achieve safety in the long term.

To have a global view of what each method is able to achieve, we depict the admissible strategies in \eqref{eq:end_admissible} and compared that with DWSC and MSNE in Figure \ref{fig:res_obj_2a}.
Both DWSC and MSNE show unsatisfactory behaviors in some types of interactions and cannot guarantee either safety or reward.
In Type A interactions, the plot matches the result in Figure \ref{fig:res_obj_1}. MSNE can achieve better safety than DWSC while DWSC generates better reward, but none of these two is optimal in either safety or reward.
In Type B interactions, DWSC reaches the safest strategy and MSNE will bring a higher reward (less than the proposed method) at the cost of safety. 
While MSNE generates the lowest rewards and the highest risks among all AAs' update rules in Type C and D interactions, DWSC achieves the highest reward there.
However, for DWSC in Type C, it will face a situation similar to Type A that DWSC will result in sub-optimal safety.   
All of these results signify the drawbacks of DWSC techniques and MSNE methods.

Then, we show that the proposed method that can control risks within a tolerable range in the long time scale and achieve a better trade-off between safety and reward than others. 
Figure \ref{fig:res_obj_2b} shows that the proposed method managed to control risk within $\epsilon$ in Type B interactions where we have no optimal value and need to make a trade-off.
As shown in Figure \ref{fig:res_obj_2a}, DWSC and MSNE only form two points that are not always optimal in all 4 types. However, in Type A and D interactions, the proposed method outperforms all others, and can even achieve the optimal strategy with the highest reward and safety. Moreover, the proposed method reaches a set of Pareto-optimal strategies, where it can make the trade-off between safety and rewards, in Type B and C interactions. Note that in Type A and C, the admissible strategy does not exist when $\epsilon$ less than a certain value, so we cannot be static there and the proposed method is actually the otimal. Although the exact shape of the plot depends on the interaction scenario's reward table and risk table, the relative relations between the proposed method and others are the same across each type.
The comparison results of different methods is summarized below.
\begin{itemize}
    \item DWSC: It can achieve the safest strategy in Type B, D and achieve the highest reward in Type C, D.
    \item MSNE: It cannot not guarantee to be the strategy with either the lowest risk or the highest reward in all types.
    \item Proposed method: In Type A and D it can achieve the optimal strategy that is both the safest and has the highest reward. In Type B and C, it can make a trade-off between the best safety and the highest reward.
\end{itemize}
All above illustrates that the proposed method is optimal and can always achieve at least the same performance as, if not better than, DWSC methods and MSNE methods.

\section{Conclusion} 
\subsubsection{Summary} 
We investigated conventional safe control methods in the presence of self-seeking humans. Our results showed that such control methods are actually not always the safest and the ones with highest performance.  Namely, we proved that when humans exploit the cooperative behaviour of autonomous agents originated from conventional safe control methods, the overall safety of the human-machine interaction decreases. Also, we proposed a policy for autonomous agents that can encourage self-seeking humans to behave in a way that optimizes safety and performance of the interaction. Moreover, our method has a better design trade-offs than existing methods like deterministic worst-case safe control, and equilibrium-based stochastic strategies.

\subsubsection{Limitations and Future Work} 
The main limitations of our work are as follows. First, we assume that the intentions of agents do not change within each episode, which is not always the case. We plan to consider intention changes within single interaction episodes in the future to further analyze the safety of interaction and improve our method. Second, the intention of agents are limited to either cooperate or defect. To extend the applicability of our method to even more general situations, we will consider other kinds of intentions, such as some more neutral intentions between the aggressive intention and conservative intention. Furthermore, it is natural that the environment changes over time and the rewards and risks may also change. Therefore, we will investigate how self-seeking humans will affect safety when we have time-varying interaction environments.

\bibliography{aaai23}

\begin{thebibliography}{36}
\providecommand{\natexlab}[1]{#1}

\bibitem[{Ahmadi et~al.(2019)Ahmadi, Singletary, Burdick, and
  Ames}]{ahmadi2019safe}
Ahmadi, M.; Singletary, A.; Burdick, J.~W.; and Ames, A.~D. 2019.
\newblock Safe policy synthesis in multi-agent POMDPs via discrete-time barrier
  functions.
\newblock In \emph{2019 IEEE 58th Conference on Decision and Control (CDC)},
  4797--4803. IEEE.

\bibitem[{Atman et~al.(2018)Atman, Hatanaka, Qu, Chopra, Yamauchi, and
  Fujita}]{atman2018motion}
Atman, M. W.~S.; Hatanaka, T.; Qu, Z.; Chopra, N.; Yamauchi, J.; and Fujita, M.
  2018.
\newblock Motion synchronization for semi-autonomous robotic swarm with a
  passivity-short human operator.
\newblock \emph{International Journal of Intelligent Robotics and
  Applications}, 2(2): 235--251.

\bibitem[{Axelrod(1984)}]{Axelrod1984-pg}
Axelrod, R. 1984.
\newblock \emph{The evolution of cooperation}.
\newblock Basic Books.

\bibitem[{Brown and Von~Neumann(1950)}]{bnn}
Brown, G.~W.; and Von~Neumann, J. 1950.
\newblock Solutions of games by differential equations.
\newblock \emph{Contributions to the Theory of Games I}, 73--79.

\bibitem[{Chen et~al.(2017)Chen, Everett, Liu, and How}]{Chen2017-ns}
Chen, Y.~F.; Everett, M.; Liu, M.; and How, J.~P. 2017.
\newblock Socially aware motion planning with deep reinforcement learning.
\newblock In \emph{2017 {IEEE/RSJ} International Conference on Intelligent
  Robots and Systems ({IROS})}, 1343--1350. IEEE.

\bibitem[{Cheng et~al.(2020)Cheng, Khojasteh, Ames, and
  Burdick}]{cheng2020safe}
Cheng, R.; Khojasteh, M.~J.; Ames, A.~D.; and Burdick, J.~W. 2020.
\newblock Safe multi-agent interaction through robust control barrier functions
  with learned uncertainties.
\newblock In \emph{2020 59th IEEE Conference on Decision and Control (CDC)},
  777--783. IEEE.

\bibitem[{Cummings et~al.(2011)Cummings, How, Whitten, and
  Toupet}]{cummings2011impact}
Cummings, M.~L.; How, J.~P.; Whitten, A.; and Toupet, O. 2011.
\newblock The impact of human--automation collaboration in decentralized
  multiple unmanned vehicle control.
\newblock \emph{Proceedings of the IEEE}, 100(3): 660--671.

\bibitem[{Dafoe et~al.(2021)Dafoe, Bachrach, Hadfield, Horvitz, Larson, and
  Graepel}]{Dafoe2021-cw}
Dafoe, A.; Bachrach, Y.; Hadfield, G.; Horvitz, E.; Larson, K.; and Graepel, T.
  2021.
\newblock Cooperative AI: machines must learn to find common ground.
\newblock \emph{Nature}, 593(7857): 33--36.

\bibitem[{Dawes(1980)}]{Dawes1980-il}
Dawes, R.~M. 1980.
\newblock Social Dilemmas.
\newblock \emph{Annual Review of Psychology}, 31(1): 169--193.

\bibitem[{Diaz-Mercado, Lee, and Egerstedt(2017)}]{diaz2017human}
Diaz-Mercado, Y.; Lee, S.~G.; and Egerstedt, M. 2017.
\newblock Human--swarm interactions via coverage of time-varying densities.
\newblock \emph{Trends in Control and Decision-Making for Human--Robot
  Collaboration Systems}, 357--385.

\bibitem[{Ding et~al.(2011)Ding, Rei{\ss}ig, Wijaya, Bortot, Bengler, and
  Stursberg}]{ding2011human}
Ding, H.; Rei{\ss}ig, G.; Wijaya, K.; Bortot, D.; Bengler, K.; and Stursberg,
  O. 2011.
\newblock Human arm motion modeling and long-term prediction for safe and
  efficient human-robot-interaction.
\newblock In \emph{2011 IEEE International Conference on Robotics and
  Automation}, 5875--5880. IEEE.

\bibitem[{Erhart and Hirche(2016)}]{erhart2016model}
Erhart, S.; and Hirche, S. 2016.
\newblock Model and analysis of the interaction dynamics in cooperative
  manipulation tasks.
\newblock \emph{IEEE Transactions on Robotics}, 32(3): 672--683.

\bibitem[{Faria, Krause, and Krause(2010)}]{Faria2010-uq}
Faria, J.~J.; Krause, S.; and Krause, J. 2010.
\newblock Collective behavior in road crossing pedestrians: the role of social
  information.
\newblock \emph{Behavioral Ecology}, 21(6): 1236--1242.

\bibitem[{Fehr and Gachter(2002)}]{Fehr2002-ur}
Fehr, E.; and Gachter, S. 2002.
\newblock Altruistic punishment in humans.
\newblock \emph{Nature}, 425: 137--140.

\bibitem[{Hilbe, Nowak, and Sigmund(2013)}]{hilbe2013evolution}
Hilbe, C.; Nowak, M.~A.; and Sigmund, K. 2013.
\newblock Evolution of extortion in iterated prisoner’s dilemma games.
\newblock \emph{Proceedings of the National Academy of Sciences}, 110(17):
  6913--6918.

\bibitem[{Hofbauer and Sigmund(1998)}]{Hofbauer1998-jg}
Hofbauer, J.; and Sigmund, K. 1998.
\newblock \emph{Evolutionary Games and Population Dynamics}.
\newblock Cambridge University Press.

\bibitem[{Ishowo-Oloko et~al.(2019)Ishowo-Oloko, Bonnefon, Soroye, Crandall,
  Rahwan, and Rahwan}]{Ishowo-Oloko2019-vn}
Ishowo-Oloko, F.; Bonnefon, J.-F.; Soroye, Z.; Crandall, J.; Rahwan, I.; and
  Rahwan, T. 2019.
\newblock Behavioural evidence for a transparency--efficiency tradeoff in
  human-machine cooperation.
\newblock \emph{Nature Machine Intelligence}, 1(11): 517--521.

\bibitem[{Kelley et~al.(2008)Kelley, Tavakkoli, King, Nicolescu, Nicolescu, and
  Bebis}]{kelley2008understanding}
Kelley, R.; Tavakkoli, A.; King, C.; Nicolescu, M.; Nicolescu, M.; and Bebis,
  G. 2008.
\newblock Understanding human intentions via hidden markov models in autonomous
  mobile robots.
\newblock In \emph{Proceedings of the 3rd ACM/IEEE international conference on
  Human robot interaction}, 367--374.

\bibitem[{Koppula and Saxena(2015)}]{koppula2015anticipating}
Koppula, H.~S.; and Saxena, A. 2015.
\newblock Anticipating human activities using object affordances for reactive
  robotic response.
\newblock \emph{IEEE transactions on pattern analysis and machine
  intelligence}, 38(1): 14--29.

\bibitem[{Kulic and Croft(2007)}]{kulic2007affective}
Kulic, D.; and Croft, E.~A. 2007.
\newblock Affective state estimation for human--robot interaction.
\newblock \emph{IEEE transactions on robotics}, 23(5): 991--1000.

\bibitem[{Luo, Sun, and Kapoor(2020)}]{luo2020multi}
Luo, W.; Sun, W.; and Kapoor, A. 2020.
\newblock Multi-robot collision avoidance under uncertainty with probabilistic
  safety barrier certificates.
\newblock \emph{Advances in Neural Information Processing Systems}, 33:
  372--383.

\bibitem[{Lyu, Luo, and Dolan(2021)}]{lyu2021probabilistic}
Lyu, Y.; Luo, W.; and Dolan, J.~M. 2021.
\newblock Probabilistic safety-assured adaptive merging control for autonomous
  vehicles.
\newblock In \emph{2021 IEEE International Conference on Robotics and
  Automation (ICRA)}, 10764--10770. IEEE.

\bibitem[{Nowak(2006)}]{Nowak2006-bl}
Nowak, M.~A. 2006.
\newblock \emph{Evolutionary Dynamics}.
\newblock Harvard University Press.

\bibitem[{Nowak and Sigmund(2005)}]{nowak2005evolution}
Nowak, M.~A.; and Sigmund, K. 2005.
\newblock Evolution of indirect reciprocity.
\newblock \emph{Nature}, 437(7063): 1291--1298.

\bibitem[{Olson(1965)}]{Olson1965-bc}
Olson, M. 1965.
\newblock \emph{The Logic of Collective Action: Public Goods and the Theory of
  Groups}.
\newblock Harvard University Press.

\bibitem[{Ostrom(1990)}]{Ostrom1990-wp}
Ostrom, E. 1990.
\newblock \emph{Governing the Commons}.
\newblock Cambridge University Press.

\bibitem[{Paiva, Santos, and Santos(2018)}]{Paiva2018-iu}
Paiva, A.; Santos, F.~P.; and Santos, F.~C. 2018.
\newblock Engineering Pro-Sociality With Autonomous Agents.
\newblock \emph{The Thirty-Second AAAI Conference on Artifical Intelligence},
  7994--7999.

\bibitem[{Peng, Carabis, and Wen(2018)}]{peng2018collaborative}
Peng, Y.-C.; Carabis, D.~S.; and Wen, J.~T. 2018.
\newblock Collaborative manipulation with multiple dual-arm robots under human
  guidance.
\newblock \emph{International Journal of Intelligent Robotics and
  Applications}, 2(2): 252--266.

\bibitem[{Rahwan et~al.(2019)Rahwan, Cebrian, Obradovich, Bongard, Bonnefon,
  Breazeal, Crandall, Christakis, Couzin, Jackson, Jennings, Kamar, Kloumann,
  Larochelle, Lazer, McElreath, Mislove, Parkes, Pentland, Roberts, Shariff,
  Tenenbaum, and Wellman}]{Rahwan2019-rk}
Rahwan, I.; Cebrian, M.; Obradovich, N.; Bongard, J.; Bonnefon, J.-F.;
  Breazeal, C.; Crandall, J.~W.; Christakis, N.~A.; Couzin, I.~D.; Jackson,
  M.~O.; Jennings, N.~R.; Kamar, E.; Kloumann, I.~M.; Larochelle, H.; Lazer,
  D.; McElreath, R.; Mislove, A.; Parkes, D.~C.; Pentland, A.~s.; Roberts,
  M.~E.; Shariff, A.; Tenenbaum, J.~B.; and Wellman, M. 2019.
\newblock Machine behaviour.
\newblock \emph{Nature}, 568(7753): 477--486.

\bibitem[{Ravichandar, Kumar, and Dani(2018)}]{ravichandar2018gaze}
Ravichandar, H.~C.; Kumar, A.; and Dani, A. 2018.
\newblock Gaze and motion information fusion for human intention inference.
\newblock \emph{International Journal of Intelligent Robotics and
  Applications}, 2(2): 136--148.

\bibitem[{Runge(1895)}]{runge1895numerische}
Runge, C. 1895.
\newblock {\"U}ber die numerische Aufl{\"o}sung von Differentialgleichungen.
\newblock \emph{Mathematische Annalen}, 46(2): 167--178.

\bibitem[{Shirado and Christakis(2020)}]{Shirado2020-hs}
Shirado, H.; and Christakis, N.~A. 2020.
\newblock Network Engineering Using Autonomous Agents Increases Cooperation in
  Human Groups".
\newblock \emph{iScience}, 23(9).

\bibitem[{Shirado, Crawford, and Christakis(2020)}]{shirado2020collective}
Shirado, H.; Crawford, F.~W.; and Christakis, N.~A. 2020.
\newblock Collective communication and behaviour in response to uncertain
  ‘Danger’in network experiments.
\newblock \emph{Proceedings of the Royal Society A}, 476(2237): 20190685.

\bibitem[{Shirado et~al.(2013)Shirado, Fu, Fowler, and
  Christakis}]{Shirado2013-nl}
Shirado, H.; Fu, F.; Fowler, J.~H.; and Christakis, N.~A. 2013.
\newblock Quality versus quantity of social ties in experimental cooperative
  networks.
\newblock \emph{Nature Communications}, 4.

\bibitem[{Smith(1984)}]{smith}
Smith, M.~J. 1984.
\newblock The stability of a dynamic model of traffic assignment—an
  application of a method of Lyapunov.
\newblock \emph{Transportation science}, 18(3): 245--252.

\bibitem[{Taylor and Jonker(1978)}]{replicator}
Taylor, P.~D.; and Jonker, L.~B. 1978.
\newblock Evolutionary stable strategies and game dynamics.
\newblock \emph{Mathematical biosciences}, 40(1-2): 145--156.

\end{thebibliography}

\appendix
\onecolumn
\section*{Appendix}
\renewcommand\qedsymbol{$\blacksquare$}

\section{Detailed Case Studies}
The details of the autonomous driving example, illustrated in Figure \ref{fig:reward_sim}, is as follows. The states of the vehicles are given by $x^i=[p^x,p^y,v,\theta]^T,\ i\in\{h,a\}$, where $p^x$ and $p^y$ are the $x$ and $y$ coordinates of the vehicle, $v$ is the speed of the vehicle, and $\theta$ is the direction the vehicle is heading to. The dynamics of the vehicles are identical and is given by
\begin{align}
    p^x_{k+1}&=p^x_k+v_k\cos(\theta_k)\Delta t\\
    p^y_{k+1}&=p^y_k+v_k\sin(\theta_k)\Delta t\\
    v_{k+1}&=v_k+(a_k+\sigma^a v_kn^a_k)\Delta t\\
    \theta_{k+1}&=\theta_k+(\phi_k+\sigma^\phi v_kn^{\phi}_k)\Delta t.
\end{align}
Here, $u_k:=[a_k,\phi_k]$ is the control input including the acceleration and the steering rate, and $\sigma^a$ and $\sigma^\phi$ determine the size of the disturbance, which depends on the current speed, in the control input. The disturbance is realized with i.i.d. $n^a_k,n^\phi_k\sim\mathcal{N}(0,1)$. The 2 vehicles must drive through a narrow road that barely lets 2 vehicles pass simultaneously. If both vehicles are defect, they pass through together with a high speed $\nu_1$. If both vehicles are cooperative, they pass through together with a reduced speed $\nu_2$. If one vehicle is cooperative and the other is defect, the defect vehicle merges into the center of the paved road and pass through with $\nu_1$, while the cooperative vehicle is forced to drive on the gravel on the side of the road with a much lower speed $\nu_3$. The disturbance makes the vehicles deviate from their desired path. Whenever a vehicle deviates enough that it enters the gravel area, it will decelerate to $\nu_3$, and when it moves back to the paved road, it will accelerate to its original speed. Acceleration costs fuel at a rate $c$. Whenever a vehicle is on the gravel and travels with a speed greater than $\nu_3$, the vehicle constantly carries a crash probability proportional to the difference between its speed and $\nu_3$, \textit{i.e.,}
\begin{align}
    \mathbb{P}(\mathcal{U}^i_k|x^i_k)\propto \max(0,v^i_k-\nu_3),
\end{align}
where $\mathcal{U}^i_k$ denotes the event that vehicle $i$ crashes at time $k$. The risk is given by the probability that any vehicle crashes at any time. The reward is given by the distance traveled by the vehicle subtracted by the fuel consumption, \textit{i.e.,}
\begin{align}
    \rho^i(x,u)=p^{xi}_K-\sum_{k=1}^{K}c\max(0,a^i_k).
\end{align}
By considering different parameters in the model, all reward and risk types can be reflected in the simulation scenario. Table~\ref{tab:all_params2} show the key parameters used to generate each types of reward and risk.

\begin{table}[h!]
  \begin{center}
    \begin{tabular}{c|c|c|c|c}
      & Type A & Type B & Type C & Type D \\
      \hline
      $\nu_1$ & 10 & 10 & 10 & 10 \\\hline
      $\nu_2$ & 6.5 & 6 & 6 & 8 \\\hline
      $\nu_3$ & 2 & 2 & 4 & 5\\\hline
      $\sigma^a$ & 0.08 & 0.15 & 0.08 & 0.2 \\\hline
      $\sigma^\phi$ & 0.08 & 0.15 & 0.08 & 0.2 \\\hline
      $c$ & 3 & 4 & 4 & 1
    \end{tabular}
    \caption{Key parameters values for different types.}
    \label{tab:all_params2}
  \end{center}
\end{table}

\section{Proofs}
As a preliminary step for the proofs; we will expand the equation of each type of human dynamics (Replicator Dynamics, Smith Dynamics, and BNN Dynamics) here. Furthermore, we will write down the mixed dynamics formula based on the three types of dynamics. Let
\begin{align}
\label{eq:alpha_def}
    \alpha = R_{CD} - R_{DD} + \pi^a_t (R_{CC}  +R_{DD}-R_{CD}-R_{DC} ).
\end{align}

\textbf{Replicator Dynamics} can be written as
\begin{align}
\dot{\pi}^h_{t} &= \pi^h_{t}(\mathbb{E}[R^h_t|\pi^a_t, \pi^h_t = 1] - \mathbb{E}[R^h_t|\pi^a_t,\pi^h_t]) \\
&= \bigg(\pi^h_{t} - (\pi^h_{t})^2\bigg)
\bigg(R_{CD} - R_{DD} + \pi^a_t (R_{CC}  +R_{DD}-R_{CD}-R_{DC} ) \bigg) \\
&= \alpha \pi^h_{t}(1 - \pi^h_{t})  \label{eq:replicator}.
\end{align}

\textbf{Smith Dynamics} can be written as (to make it more readable, we redefine $\mathbb{E}[R^h_t|\pi^a_t, \pi^h_t = 1]$ as $\theta_C^h(t)$, and $\mathbb{E}[R^h_t|\pi^a_t, \pi^h_t = 0]$ as $\theta_D^h(t)$)
\begin{align}
\dot{\pi}^h_{t} &= (1-\pi^h_{t}) \bigg[\theta_C^h(t) - \theta_D^h(t) \bigg]_{+} - \pi^h_{t}\bigg[\theta_D^h(t) - \theta_C^h(t)  \bigg]_{+} \label{eq:smith_1} \\
&= 
\begin{cases}		 
	    (1-\pi^h_{t}) \bigg(\theta_C^h(t) - \theta_D^h(t) \bigg), & \theta_C^h(t) > \theta_D^h(t)\\
	    - \pi^h_{t}\bigg(\theta_D^h(t) - \theta_C^h(t)  \bigg), & \theta_C^h(t) < \theta_D^h(t)  \\
	    0, & \theta_C^h(t) = \theta_D^h(t)
\end{cases}	\label{eq:smith_2} \\ 
&=
\begin{cases}		 
	    \alpha (1-\pi_t^h), & \pi^a_t < L\\
	    \alpha \pi_t^h, &  \pi^a_t > L  \\
	    0, &\pi^a_t = L.
\end{cases}	 \label{eq:smith_4}
\end{align}
Here, (\ref{eq:smith_4}) conditions is due to the following relationship under the assumptions \eqref{eq:assumption2}--\eqref{eq:assumption3}:
\begin{align}
\label{eq:c_d_rel}
\theta_C^h(t) - \theta_D^h(t) =  R_{CD} - R_{DD} + \pi^a_t (R_{CC} +R_{DD}-R_{CD}-R_{DC}).
\end{align}

\textbf{BNN Dynamics} can be written as (to make it more readable, we redefine $\mathbb{E}[R^h_t|\pi^a_t, \pi^h_t = 1]$ as $\theta_C^h(t)$, $\mathbb{E}[R^h_t|\pi^a_t, \pi^h_t = 0]$ as $\theta_D^h(t)$, and $\mathbb{E}[R^h_t|\pi^a_t,\pi^h_t]$ as $\bar{\theta}^h(t)$)
\begin{align}
\dot{\pi}^h_{t} &= 
 \bigg[\theta_C^h(t) - \bar{\theta}^h(t) \bigg]_{+} - \pi^h_{t} \bigg( \bigg[\theta_C^h(t) - \bar{\theta}^h(t) \bigg]_{+} + \bigg[\theta_D^h(t) - \bar{\theta}^h(t) \bigg]_{+} \bigg) \label{eq:bnn_1}\\
 &= \begin{cases}
	      \theta_C^h(t) - \bar{\theta}^h(t) - \pi^h_t \bigg(\theta_C^h(t) - \bar{\theta}^h(t)\bigg),  & \theta_C^h(t) > \bar{\theta}^h(t) \geq \theta_D^h(t) \\
	       - \pi^h_t \bigg(\theta_D^h(t) - \bar{\theta}^h(t)\bigg), & \theta_D^h(t) > \bar{\theta}^h(t) \geq \theta_C^h(t) \\
	      0,  &  \theta_C^h(t) = \bar{\theta}^h(t) > \theta_D^h(t) \\
	       0, &  \theta_D^h(t) = \bar{\theta}^h(t) > \theta_C^h(t) \\	       
	       0, &  \theta_D^h(t) = \bar{\theta}^h(t) = \theta_C^h(t) \\
    \end{cases} \label{eq:bnn_2} \\
&= \left\{
	\begin{array}{ll}
	       \alpha (1-\pi_t^h)^2, & \pi^a_t < L \text{ and } \pi_t^h \in [0,1) \\
	       \alpha (\pi_t^h)^2, &  \pi^a_t > L \text{ and } \pi_t^h \in (0,1]  \\
	       0,  &  \pi^a_t < L \text{ and } \pi^h_t = 1 \\
	       0, &  \pi^a_t > L \text{ and } \pi^h_t = 0 \\
	       0, & \pi^a_t = L.\\
	\end{array}
\right. \label{eq:bnn_3}
\end{align}
Here, (\ref{eq:bnn_3}) conditions is due to the following relationships under the assumptions \eqref{eq:assumption2}--\eqref{eq:assumption3}.
\begin{gather}
    \theta_C^h(t) - \theta_D^h(t) =  R_{CD} - R_{DD} + \pi^a_t (R_{CC} +R_{DD}-R_{CD}-R_{DC}) \\
\bar{\theta}^h(t) = \pi^h_t \theta_C^h(t) + (1 - \pi^h_t) \theta_D^h(t).
\end{gather}

\textbf{Mixed Dynamics} is written as a weighted sum of the three types of dynamics discussed earlier as follows
\begin{align}
     \dot{\pi}^h_t &=
	\begin{cases}
	       \alpha \bigg(w_r \pi^h_t(1-\pi^h_t) + w_s (1-\pi^h_t) + w_b (1-\pi_t^h)^2  \bigg) ,& \pi^a_t < L \text{ and } \pi_t^h \in [0,1) \\
		   \alpha \bigg(w_r \pi^h_t(1-\pi^h_t) + w_s (1-\pi^h_t) \bigg) ,& \pi^a_t < L \text{ and } \pi^h_t = 1 \\
	       \alpha \bigg(w_r \pi^h_t(1-\pi^h_t) + w_s (\pi^h_t) + w_b (\pi^h_t)^2  \bigg) ,&  \pi^a_t > L \text{ and } \pi_t^h \in (0,1]  \\
	       \alpha \bigg(w_r \pi^h_t(1-\pi^h_t) + w_s (\pi^h_t) \bigg) ,& \pi^a_t > L \text{ and } \pi^h_t = 0 \\
	       \alpha \bigg(w_r \pi^h_t(1-\pi^h_t) \bigg) ,&\pi^a_t = L,
	\end{cases} \label{eq:mixed_1}
\end{align}
where $w_r, w_s, w_b$ are the positive weights of each type of dynamics.

\begin{proof}[\textbf{Proof of Lemma \ref{lem:dwsc_not_safe}}]

Due to \eqref{eq:assumption2}, \eqref{eq:assumption3}, and \eqref{eq:assumption1}, when $\pi^a_t=1 > L$, we have
\begin{align}
    \alpha=R_{CC}-R_{DC}<0.
\end{align}
Due to \eqref{eq:mixed_1}, when $\pi^a_t > L \text{ and } \pi_t^h \in (0,1]$, we have
\begin{align}
    \dot{\pi}^h_t=\alpha \bigg(w_r \pi^h_t(1-\pi^h_t) + w_s (\pi^h_t) + w_b (\pi^h_t)^2  \bigg).
\end{align}
Since 
\begin{align}
    \alpha<0,
\end{align}
and
\begin{align}
    w_r \pi^h_t(1-\pi^h_t) + w_s \pi^h_t + w_b (\pi^h_t)^2>0,\ \forall \pi^h_t\in(0,1],
\end{align}
we have
\begin{align}
    \dot{\pi}^h_t<0,\ \forall t\geq 0.\label{eq:corr_result1}
\end{align}
Which means $\pi^h_t$ is monotonically decreasing, so
\begin{align}
    \pi^h_t\xrightarrow[]{t\rightarrow\infty} 0.
\end{align}
Whereas, due to \eqref{eq:mixed_1}, when $\pi^a_t > L \text{ and } \pi^h_t = 0$, we have
\begin{align}
    \dot{\pi}^h_t&=\alpha \bigg(w_r \pi^h_t(1-\pi^h_t) + w_s (\pi^h_t)\bigg) \\
    &= 0.
\end{align}
\end{proof}

\begin{proof}[\textbf{Proof of Lemma \ref{lem:stop}}]
We recall the admissible strategy subsets
\begin{align}
    \mathcal{A}_0&=\{(\pi^h,\pi^a):\pi^h=0,\pi^a\in(L,1]\}\label{eq:A0}\\
    \mathcal{A}_1&=\{(\pi^h,\pi^a):\pi^h=1,\pi^a\in[0,L)\} \label{eq:A1}\\
    \mathcal{A}_d&=\{(\pi^h,\pi^a):\pi^h\in[0,1],\pi^a=L\} \label{eq:AD},
\end{align}
where the set of admissible strategy $\mathcal{A}$ is given by
\begin{align}
\label{eq:end_admissible_appendix}
    \mathcal{A}=\mathcal{A}_0\cup\mathcal{A}_1\cup\mathcal{A}_d.
\end{align}
From \eqref{eq:mixed_1}, we see that $\mathcal{A}_0$ corresponds to the fourth case condition. Whereas, $\mathcal{A}_1$ corresponds to the second case condition. Lastly, $\mathcal{A}_d$ corresponds to the last case condition. Accordingly, we get the following
\begin{align}
     \dot{\pi}^h_t &=
	\begin{cases}
	       \alpha \bigg(w_r \pi^h_t(1-\pi^h_t) + w_s (\pi^h_t) \bigg) ,& \pi_t \in \mathcal{A}_0 \\
		   \alpha \bigg(w_r \pi^h_t(1-\pi^h_t) + w_s (1-\pi^h_t) \bigg) ,& \pi_t \in \mathcal{A}_1  \\
	       \alpha \bigg(w_r \pi^h_t(1-\pi^h_t) \bigg) ,& \pi_t \in \mathcal{A}_d.
	\end{cases} \label{eq:l1_1}
\end{align}
Therefore, from \eqref{eq:A0}--\eqref{eq:AD}, we have
\begin{align}
    \textrm{First case: } \pi_t \in \mathcal{A}_0 &\Rightarrow \pi^h_t=0 \Rightarrow \dot{\pi}^h_t = 0 \label{eq:l3_step_1_1}\\
    \textrm{Second case: } \pi_t \in \mathcal{A}_1  &\Rightarrow \pi^h_t=1 \Rightarrow \dot{\pi}^h_t = 0 \label{eq:l3_step_1_2}\\
    \textrm{Third case: } \pi_t \in \mathcal{A}_d &\Rightarrow \pi^a_t=L.
    \label{eq:lm1_temp_1}
\end{align}

Given \eqref{eq:alpha_def}, we have
\begin{align}
    \pi^a_t=\frac{\alpha-R_{CD}+R_{DD}}{R_{CC}+R_{DD}-R_{CD}-R_{DC}}.
\end{align}
Therefore, \eqref{eq:lm1_temp_1} continues as follows
\begin{align}
    \pi^a_t=L&\Rightarrow\frac{\alpha-R_{CD}+R_{DD}}{R_{CC}+R_{DD}-R_{CD}-R_{DC}}=\frac{R_{DD}-R_{CD}}{R_{CC}+R_{DD}-R_{DC}-R_{CD}} \Rightarrow \alpha=0 \Rightarrow \dot{\pi}^h_t = 0.
    \label{eq:lm1_temp_1_end}
\end{align}
    
To show the other direction of the proof, we have to show that $\pi_t \notin \mathcal{A} \Rightarrow \dot{\pi}^h_t \neq 0$. The two remaining cases from \eqref{eq:mixed_1} that is not in \eqref{eq:l1_1} are given by
\begin{align}
     \dot{\pi}^h_t &=
	\begin{cases}
	       \alpha \bigg(w_r \pi^h_t(1-\pi^h_t) + w_s (1-\pi^h_t) + w_b (1-\pi_t^h)^2  \bigg) ,& \pi^a_t < L \text{ and } \pi_t^h \in [0,1) \\
	       \alpha \bigg(w_r \pi^h_t(1-\pi^h_t) + w_s (\pi^h_t) + w_b (\pi^h_t)^2  \bigg) ,&  \pi^a_t > L \text{ and } \pi_t^h \in (0,1].  \\
	\end{cases} \label{eq:other_direction1}
\end{align}
Therefore, we have
\begin{align}
    \textrm{First case: } &\pi^a_t < L \textrm{ and } \pi_t^h \in [0,1) \Rightarrow \alpha \neq 0 \Rightarrow \dot{\pi}^h_t \neq 0 \label{eq:other_direction2}\\
    \textrm{Second case: } &\pi^a_t > L \textrm{ and } \pi_t^h \in (0,1] \Rightarrow \alpha \neq 0 \Rightarrow \dot{\pi}^h_t \neq 0. \label{eq:other_direction3}
\end{align}

\end{proof}
\begin{proof}[\textbf{Proof of Theorem 1}]
We recall our proposed policy \begin{align}
\label{eq:proposed_policy}
    \pi^a_t=\begin{cases}
        L-G(\pi^{h \star}-\pi^h_t), & \pi^{\star}\in\mathcal{F}_d\\
        \pi^{a \star}, & \textrm{otherwise},
    \end{cases}
\end{align}
where $G \in \mathbb R$ is a strictly positive constant. We first show that
\begin{align}
\label{eq:alpha_small}
    \pi^a_t=\pi^{a \star}>L\Rightarrow\alpha<0
\end{align}
and
\begin{align}
\label{eq:alpha_large}
    \pi^a_t=\pi^{a \star}<L\Rightarrow\alpha>0.
\end{align}
Given \eqref{eq:alpha_def}, we have
\begin{align}
    \pi^a_t=\frac{\alpha-R_{CD}+R_{DD}}{R_{CC}+R_{DD}-R_{CD}-R_{DC}}.
\end{align}
Therefore,
\begin{align}
    \pi^a_t>L&\Rightarrow\frac{\alpha-R_{CD}+R_{DD}}{R_{CC}+R_{DD}-R_{CD}-R_{DC}}>\frac{R_{DD}-R_{CD}}{R_{CC}+R_{DD}-R_{DC}-R_{CD}} \label{eq:alpha_leq_z1}\\
    \pi^a_t<L&\Rightarrow\frac{\alpha-R_{CD}+R_{DD}}{R_{CC}+R_{DD}-R_{CD}-R_{DC}}<\frac{R_{DD}-R_{CD}}{R_{CC}+R_{DD}-R_{DC}-R_{CD}} \label{eq:alpha_geq_z1}.
\end{align}
Due to \eqref{eq:assumption2}--\eqref{eq:assumption3}, we have
\begin{align}
    \frac{\alpha-R_{CD}+R_{DD}}{R_{CC}+R_{DD}-R_{CD}-R_{DC}}>\frac{R_{DD}-R_{CD}}{R_{CC}+R_{DD}-R_{DC}-R_{CD}}&\Rightarrow\alpha<0 \label{eq:alpha_leq_z2} \\
    \frac{\alpha-R_{CD}+R_{DD}}{R_{CC}+R_{DD}-R_{CD}-R_{DC}}<\frac{R_{DD}-R_{CD}}{R_{CC}+R_{DD}-R_{DC}-R_{CD}}&\Rightarrow\alpha>0 \label{eq:alpha_geq_z2}.
\end{align}
Thus, we show \eqref{eq:alpha_small} and \eqref{eq:alpha_large}. Next, we show that when $\pi^\star\in\mathcal{F}_0$, we have
\begin{align}
    \pi^a_t=\pi^{a \star}, \forall t\geq 0\Rightarrow\pi^h_t\xrightarrow[]{t\rightarrow\infty}\pi^{h \star} = 0,\ \forall \pi^h_0\in(0,1).
\end{align}
Here, the interval for $\pi^h_0$ is based on the assumption that $\pi^h_0\neq 0$ and $\pi^h_0\neq 1$. From the definition of $\mathcal{F}_0$, we have
\begin{align}
    \pi^a_t = \pi^{a \star}&>L, \forall t\geq 0,\label{eq:thm1_temp_1}
\end{align}
which, due to \eqref{eq:mixed_1}, gives
\begin{align}
    \dot{\pi}^h_t=\alpha \bigg(w_r \pi^h_t(1-\pi^h_t) + w_s (\pi^h_t) + w_b (\pi^h_t)^2  \bigg),
\end{align}
and, due to \eqref{eq:alpha_small}, gives
\begin{align}
    \alpha<0.
\end{align}
Since
\begin{align}
    w_r \pi^h_t(1-\pi^h_t) + w_s \pi^h_t + w_b (\pi^h_t)^2>0,\ \forall \pi^h_t\in(0,1),
\end{align}
we have
\begin{align}
    \dot{\pi}^h_t<0,\ \forall t\geq 0,\label{eq:thm1_temp_2}
\end{align}
which means $\pi^h_t$ is monotonically decreasing whenever $\pi^a_t>L$. Since $\pi^h_t\geq 0$, we have
\begin{align}\label{eq:greater_than_L}
    \pi^h_t\xrightarrow[]{t\rightarrow\infty}\pi^{h \star}=0.
\end{align}
Similarly, we show that when $\pi^\star\in\mathcal{F}_1$, we have
\begin{align}
    \pi^a_t=\pi^{a \star}, \forall t\geq 0\Rightarrow\pi^h_t\xrightarrow[]{t\rightarrow\infty}\pi^{h \star} = 1 ,\ \forall \pi^h_0\in(0,1).
\end{align}
From the definition of $\mathcal{F}_1$, we have
\begin{align}
    \pi^a_t = \pi^{a \star}&<L, \forall t\geq 0, \label{eq:thm1_temp_3}
\end{align}
which, due to \eqref{eq:mixed_1}, gives
\begin{align}
    \dot{\pi}^h_t=\alpha \bigg(w_r \pi^h_t(1-\pi^h_t) + w_s (1-\pi^h_t) + w_b (1-\pi^h_t)^2  \bigg),
\end{align}
and, due to \eqref{eq:alpha_large}, gives
\begin{align}
    \alpha>0.
\end{align}
Since
\begin{align}
    w_r \pi^h_t(1-\pi^h_t) + w_s (1-\pi^h_t) + w_b (1-\pi^h_t)^2>0,\ \forall \pi^h_t\in(0,1),
\end{align}
we have
\begin{align}
    \dot{\pi}^h_t>0,\ \forall t\geq 0,\label{eq:thm1_temp_4}
\end{align}
which means $\pi^h_t$ is monotonically increasing whenever $\pi^a_t<L$. Since $\pi^h_t\leq 1$, we have
\begin{align}\label{eq:less_than_L}
    \pi^h_t\xrightarrow[]{t\rightarrow\infty}\pi^{h \star}=1.
\end{align}
Next, we show that, when $\pi^\star\in\mathcal{F}_d$, we have
\begin{align}
    \pi^a_t=L-G(\pi^{h \star}-\pi^h_t), \forall t\geq 0\Rightarrow\pi^h_t\xrightarrow[]{t\rightarrow\infty}\pi^{h \star},\ \forall \pi^h_0\in(0,1).
\end{align}
From the definition of $\mathcal{F}_d$, we have
\begin{align}
    \pi^{a \star}&=L.
\end{align}
Since $G>0$, we have
\begin{align}
\pi^h_t=\pi^{h \star}&\Rightarrow\pi^a_t=L\Rightarrow\dot{\pi}^h_t=0 \label{eq:thm1_fd1}
\end{align}
from \eqref{eq:lm1_temp_1}--\eqref{eq:lm1_temp_1_end}. In the other case
\begin{align}
\pi^h_t>\pi^{h \star}&\Rightarrow\pi^a_t>L,
\end{align}
which, due to \eqref{eq:mixed_1}, gives
\begin{align}
    \dot{\pi}^h_t=\alpha \bigg(w_r \pi^h_t(1-\pi^h_t) + w_s (\pi^h_t) + w_b (\pi^h_t)^2  \bigg),
\end{align}
and, due to \eqref{eq:alpha_small}, gives
\begin{align}
    \alpha<0.
\end{align}
Since
\begin{align}
    w_r \pi^h_t(1-\pi^h_t) + w_s \pi^h_t + w_b (\pi^h_t)^2>0,\ \forall \pi^h_t\in(0,1),
\end{align}
we have
\begin{align}
    \dot{\pi}^h_t<0,\ \forall t\geq 0,
\end{align}
with this, we have shown
\begin{align}
\pi^h_t>\pi^{h \star}&\Rightarrow\pi^a_t>L\Rightarrow\dot{\pi}^h_t<0. \label{eq:thm1_fd2}
\end{align}
In the last case, we have
\begin{align}
\pi^h_t<\pi^{h \star}&\Rightarrow\pi^a_t<L,
\end{align}
which, due to \eqref{eq:mixed_1}, gives
\begin{align}
    \dot{\pi}^h_t=\alpha \bigg(w_r \pi^h_t(1-\pi^h_t) + w_s (1-\pi^h_t) + w_b (1-\pi^h_t)^2  \bigg),
\end{align}
and, due to \eqref{eq:alpha_large}, gives
\begin{align}
    \alpha>0.
\end{align}
Since
\begin{align}
    w_r \pi^h_t(1-\pi^h_t) + w_s (1-\pi^h_t) + w_b (1-\pi^h_t)^2>0,\ \forall \pi^h_t\in(0,1),
\end{align}
we have
\begin{align}
    \dot{\pi}^h_t>0,\ \forall t\geq 0,
\end{align}
with this, we have shown
\begin{align}
\pi^h_t<\pi^{h \star}&\Rightarrow\pi^a_t<L\Rightarrow\dot{\pi}^h_t>0. \label{eq:thm1_fd3}
\end{align}
From \eqref{eq:thm1_fd1}, \eqref{eq:thm1_fd2}, and \eqref{eq:thm1_fd3}, $\pi^h_t$ will eventually converge to $\pi^{h \star}$, \textit{i.e.,}
\begin{align}
    \pi^h_t\xrightarrow[]{t\rightarrow\infty}\pi^{h \star},
\end{align}
at which time,
\begin{align}
    \pi^a_t=\pi^{a \star}=L.
\end{align}
Finally, we show that
\begin{align}
\label{eq:f_safe}
    \pi=(\pi^h,\pi^a)\in\mathcal{F}\Rightarrow\mathbb{E}[W|\pi^h,\pi^a]\leq\epsilon,
\end{align}
where
\begin{align}
    \mathbb{E}[W|\pi^h,\pi^a]&=W_{CC}\pi^h\pi^a+W_{CD}\pi^h(1-\pi^a)+W_{DC}(1-\pi^h)\pi^a+W_{DD}(1-\pi^h)(1-\pi^a)\\
    &=W_{CC}\pi^h\pi^a+W_{CD}\pi^h(1-\pi^a)+W_{CD}(1-\pi^h)\pi^a+W_{DD}(1-\pi^h)(1-\pi^a), \label{eq:use_cd_dc}
\end{align}
where \eqref{eq:use_cd_dc} is due to the assumption that $W_{CD}=W_{DC}$. We start by showing that
\begin{align}
\label{eq:f0_safe}
    \pi=(\pi^h,\pi^a)\in\mathcal{F}_0\Rightarrow\mathbb{E}[W|\pi^h,\pi^a]\leq\epsilon.
\end{align}
Plugging $\pi^h=0$ into \eqref{eq:use_cd_dc} gives
\begin{align}
    \mathbb{E}[W|\pi^h,\pi^a]=W_{CD}\pi^a+W_{DD}(1-\pi^a)\leq\epsilon,
\end{align}
which yields
\begin{align}
    (W_{CD}-W_{DD})\pi^a\leq\epsilon-W_{DD}.
\end{align}
Due to the assumption that $W_{CD}<W_{DD}$, we have
\begin{align}
    \pi^a\geq \frac{\epsilon-W_{DD}}{W_{CD}-W_{DD}},
\end{align}
which implies, when $\pi^h=0$,
\begin{align}
    \pi^a\in\mathcal{B}_0\Rightarrow\mathbb{E}[W|\pi^h,\pi^a]\leq\epsilon,
\end{align}
which further implies \eqref{eq:f0_safe}. Similarly, we show that
\begin{align}
\label{eq:f1_safe}
    \pi=(\pi^h,\pi^a)\in\mathcal{F}_1\Rightarrow\mathbb{E}[W|\pi^h,\pi^a]\leq\epsilon.
\end{align}
Plugging $\pi^h=1$ into \eqref{eq:use_cd_dc} gives
\begin{align}
    \mathbb{E}[W|\pi^h,\pi^a]=W_{CC}\pi^a+W_{CD}(1-\pi^a)\leq\epsilon,
\end{align}
which yields
\begin{align}
    (W_{CC}-W_{CD})\pi^a\leq\epsilon-W_{CD}.
\end{align}
Rearranging yields
\begin{align}
    \begin{cases}
        \pi^a\leq\frac{\epsilon-W_{CD}}{W_{CC}-W_{CD}}, & W_{CC}>W_{CD} \\ 
        \pi^a\geq\frac{\epsilon-W_{CD}}{W_{CC}-W_{CD}}, & W_{CC}<W_{CD},
    \end{cases}
\end{align}
which implies, when $\pi^h=1$,
\begin{align}
    \pi^a\in\mathcal{B}_1\Rightarrow\mathbb{E}[W|\pi^h,\pi^a]\leq\epsilon,
\end{align}
which further implies \eqref{eq:f1_safe}. Lastly, we show that
\begin{align}
\label{eq:fd_safe}
    \pi=(\pi^h,\pi^a)\in\mathcal{F}_d\Rightarrow\mathbb{E}[W|\pi^h,\pi^a]\leq\epsilon.
\end{align}
Plugging $\pi^a=L$ into \eqref{eq:use_cd_dc} gives
\begin{align}
    \mathbb{E}[W|\pi^h,\pi^a]=W_{CC}\pi^hL+W_{CD}\pi^h(1-L)+W_{CD}(1-\pi^h)L+W_{DD}(1-\pi^h)(1-L)\leq\epsilon,
\end{align}
which yields
\begin{align}
    (W_{CC}L+W_{CD}-2W_{CD}L-W_{DD}(1-L))\pi^h<\epsilon-W_{CD}L-W_{DD}(1-L).
\end{align}
Rearranging yields
\begin{align}
    \begin{cases}
        \pi^h\leq\frac{\epsilon-W_{CD}L-W_{DD}(1-L)}{W_{CC}L+W_{CD}-2W_{CD}L-W_{DD}(1-L)}, & W_{CC}L+W_{CD}-2W_{CD}L-W_{DD}(1-L)>0 \\
        \pi^h\geq\frac{\epsilon-W_{CD}L-W_{DD}(1-L)}{W_{CC}L+W_{CD}-2W_{CD}L-W_{DD}(1-L)}, & W_{CC}L+W_{CD}-2W_{CD}L-W_{DD}(1-L)<0,
    \end{cases}
\end{align}
which implies, when $\pi^a=L$,
\begin{align}
    \pi^h\in\mathcal{B}_d\Rightarrow\mathbb{E}[W|\pi^h,\pi^a]\leq\epsilon,
\end{align}
which further implies \eqref{eq:fd_safe}. Since
\begin{align}
    \mathcal{F}=\mathcal{F}_0\cup \mathcal{F}_1\cup \mathcal{F}_d,
\end{align}
we prove \eqref{eq:f_safe}.
\end{proof}

\section{Additional Numerical Experiments}

\begin{figure}[H]
\begin{subfigure}{0.5\textwidth}
    \centering
    \includegraphics[width=1\linewidth]{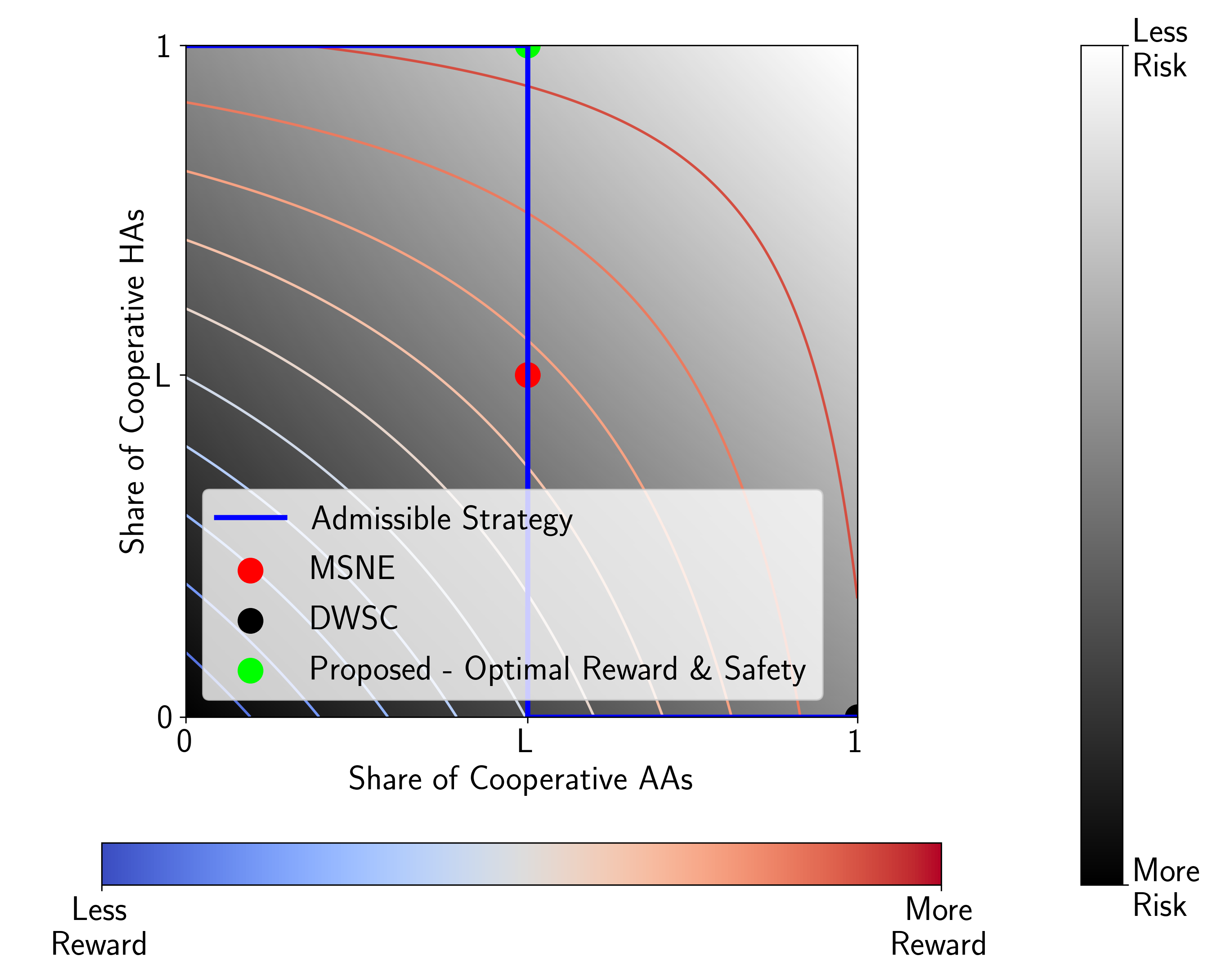}
    \caption{Type A}
    \label{fig:risk_map_type_A}
\end{subfigure}%
\begin{subfigure}{0.5\textwidth}
    \centering
    \includegraphics[width=1\linewidth]{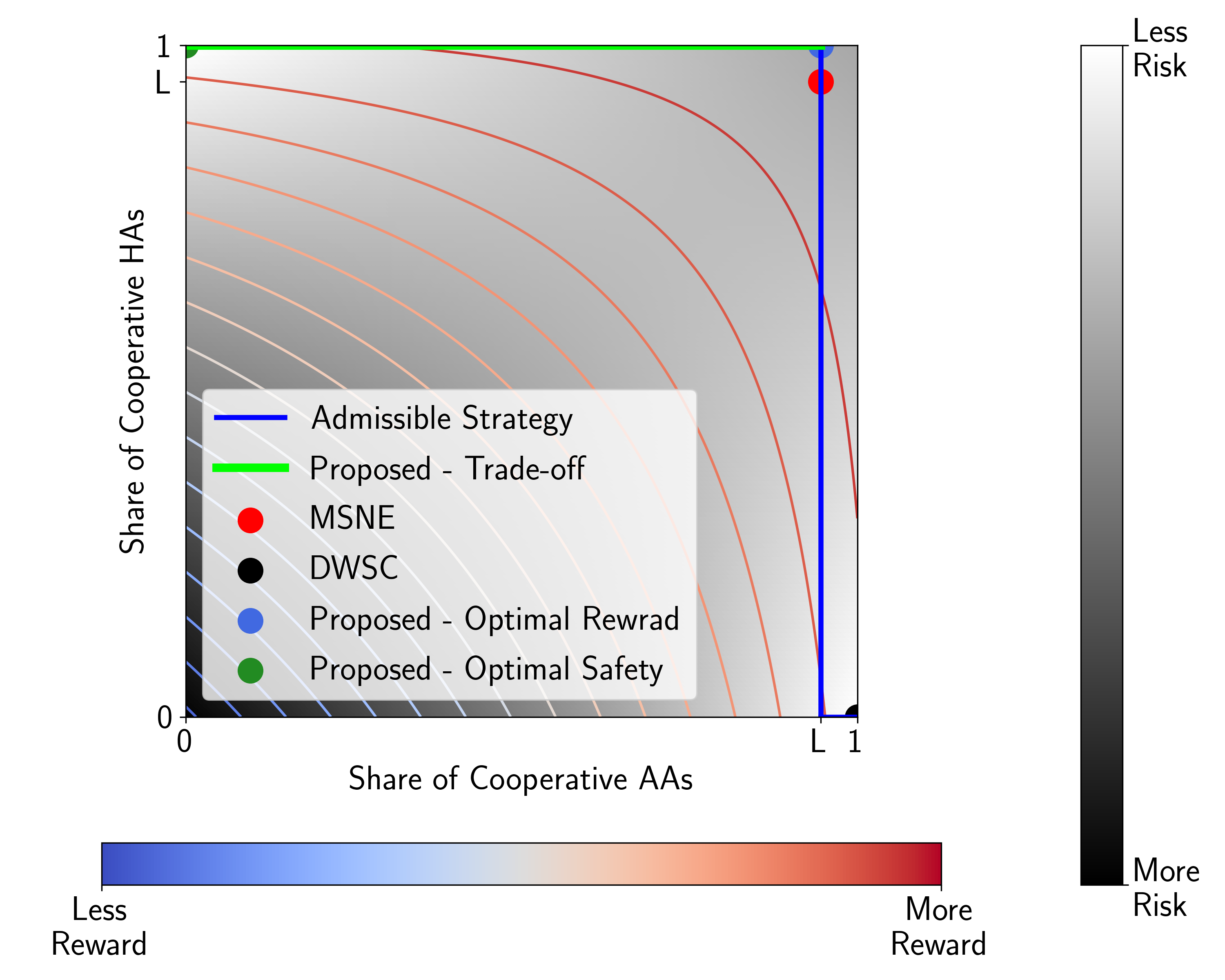}
    \caption{Type B}
    \label{fig:risk_map_type_B}
\end{subfigure}%
\\
\begin{subfigure}{0.5\textwidth}
    \centering
    \includegraphics[width=1\linewidth]{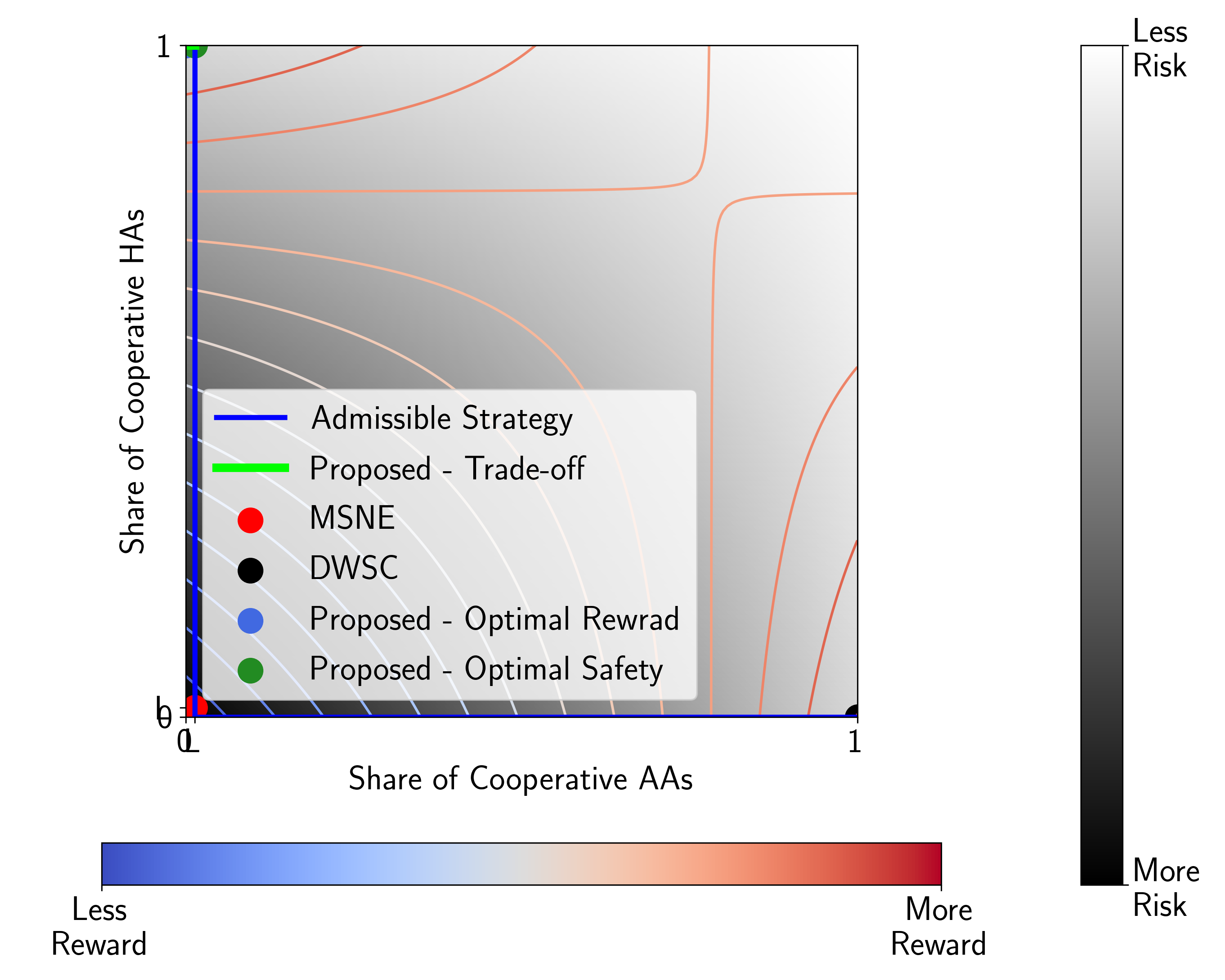}
    \caption{Type C}
    \label{fig:risk_map_type_C}
\end{subfigure}%
\begin{subfigure}{0.5\textwidth}
    \centering
    \includegraphics[width=1\linewidth]{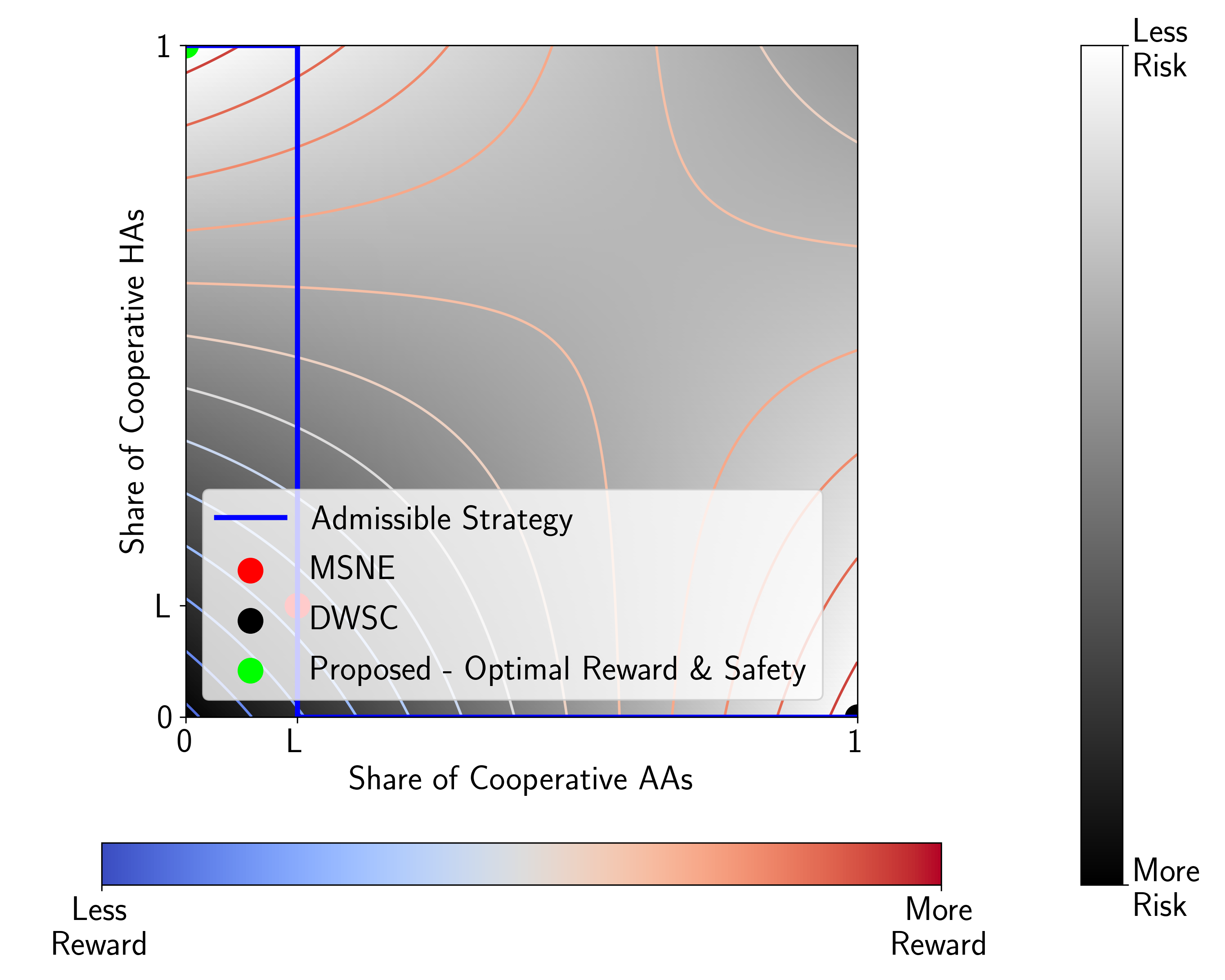}
    \caption{Type D}
    \label{fig:risk_map_type_D}
\end{subfigure}%
\caption{Risk maps with contours indicating the reward levels.}
\end{figure}
We start by showing the intuitive comparison of the strategies achievable by MSNE, DWSC, and our proposed method. Figure \ref{fig:risk_map_type_A} shows the risk map for Type A interactions. Since, in this type, the regions with the highest expected total rewards are also the regions with the lowest expected risk, the strategy that maximizes the expected total reward also minimizes the risk. We can see that the proposed strategy has an advantage over both MSNE and DWSC in terms of both reward maximization and risk minimization. 

Figure \ref{fig:risk_map_type_B} shows the risk map for Type B interactions. Since, in this type, the regions with the highest expected total reward is not the regions with the lowest expected risk, there exists a tradeoff between performance and safety. We show the proposed method in its tradeoff region has an advantage over MSNE in terms of both reward maximization and risk minimization. Although the proposed method, when minimizing risk, yields the strategy that has the same performance and safety as DWSC due to symmetry, DWSC can only achieve this strategy. On the other hand, the proposed method can achieve different strategies based on different safety specifications and find the maximum reward possible.

Figure \ref{fig:risk_map_type_C} shows the risk map for Type C interactions. Similar to Type B, there also exists a tradeoff between performance and safety, and the proposed method in its tradeoff region also has an advantage over MSNE in terms of both reward maximization and risk minimization. The proposed method, when maximizing reward, yields the strategy that has the same performance and safety as DWSC due to symmetry. Similar to Type B, DWSC can only achieve this strategy, but the proposed method can achieve different strategies based on different safety specifications and find the maximum reward possible.

Figure \ref{fig:risk_map_type_D} shows the risk map for Type D interactions. Similar to Type A, the strategy that maximizes the expected total reward also minimizes the risk. In this case, DWSC and the proposed method both find the optimal strategy. In this case, the proposed method does not have an advantage over DWSC.


\begin{figure}[h]
    \centering
    \includegraphics[width=0.6\linewidth]{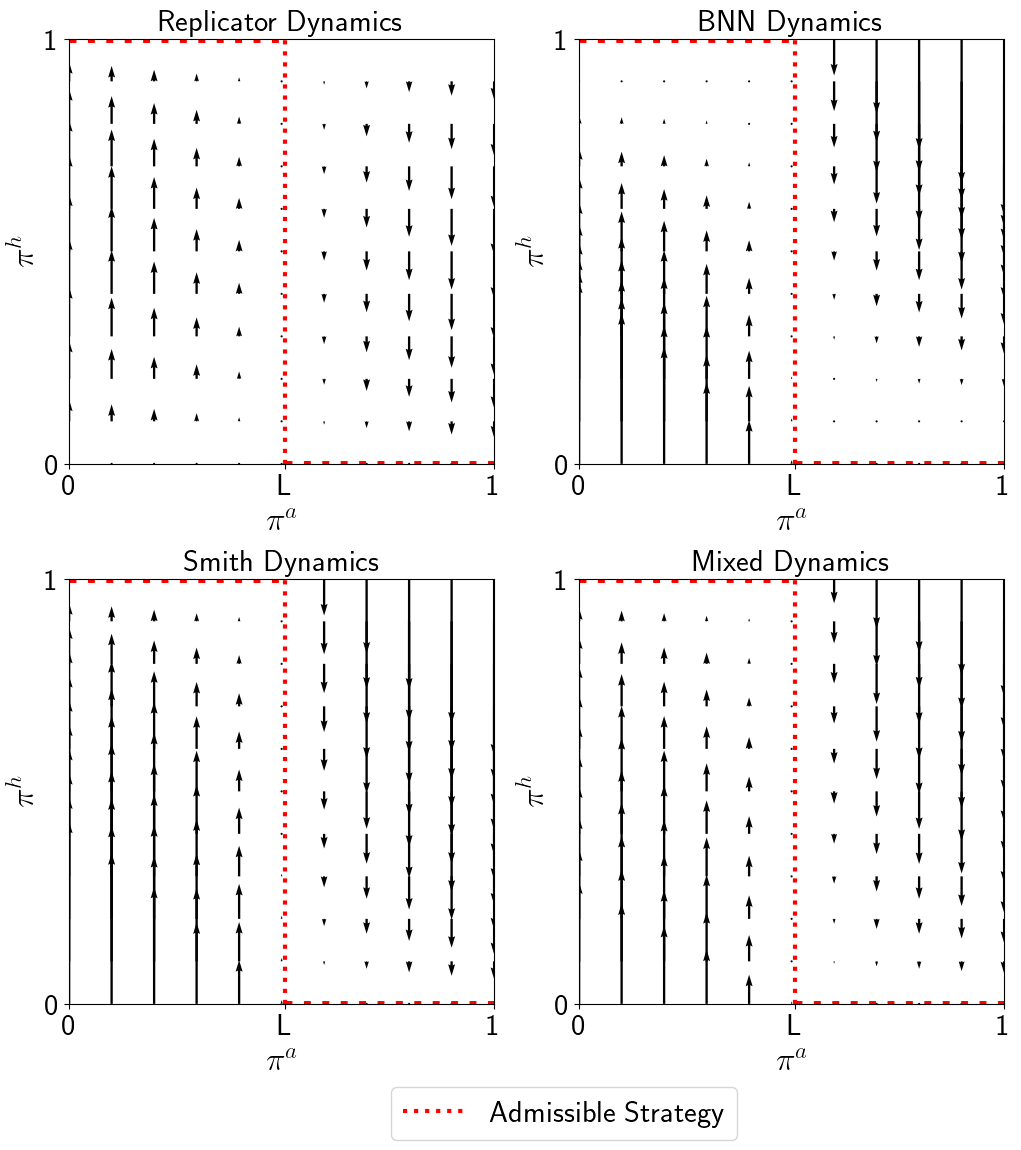}
    \caption{Admissible strategy of Type A interaction.}
    \label{fig:admissible_strategy_type_A}
\end{figure}

\begin{figure}[h]
    \centering
    \includegraphics[width=0.6\linewidth]{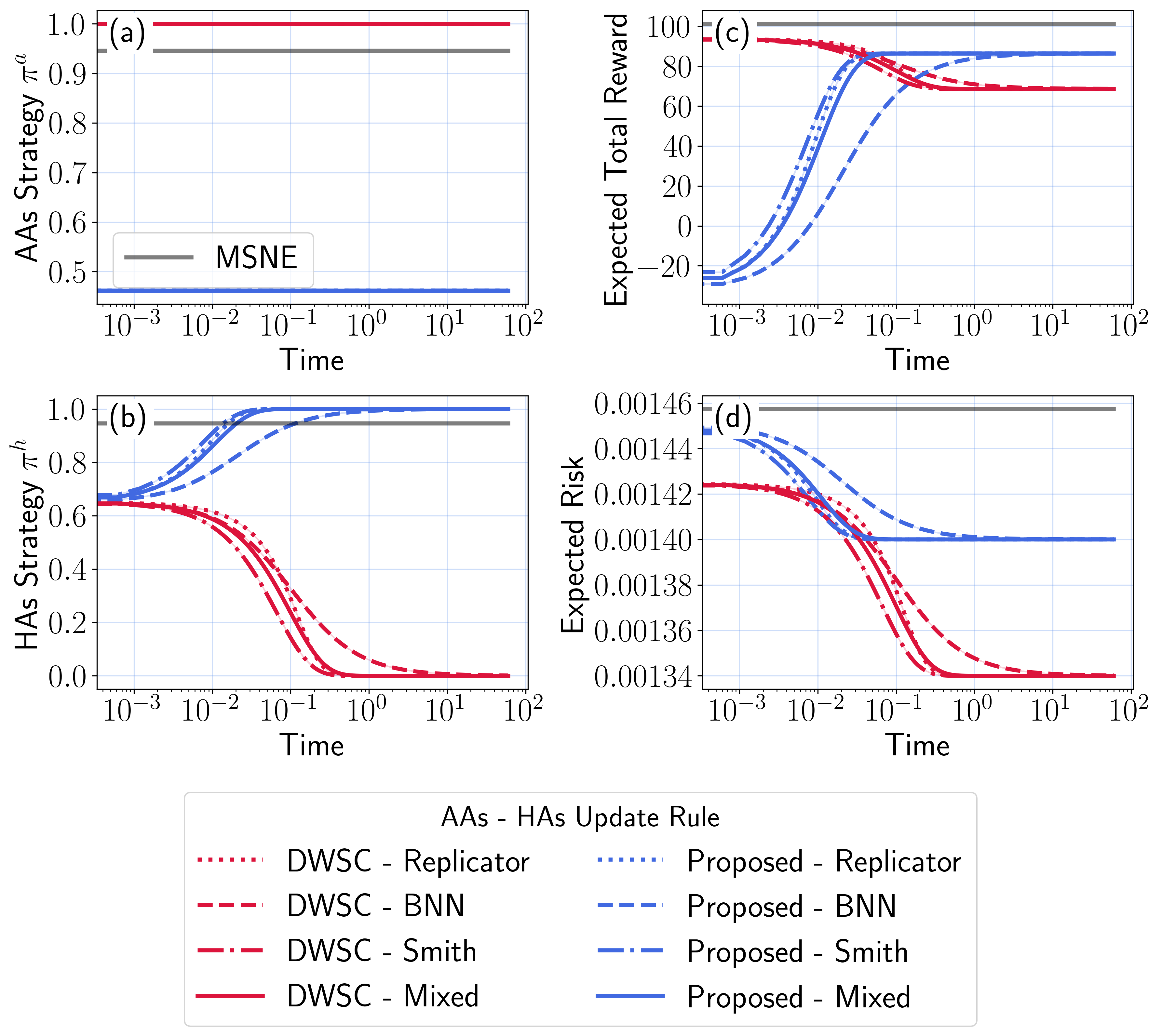}
    \caption{Evolution of states in Type B interaction with different AAs-HAs Update Rules ($\pi^h(0)=0.65$ and $\epsilon=1.4e-3$).}
    \label{fig:res_type_B}
\end{figure}
\begin{figure}[h]
    \centering
    \includegraphics[width=0.6\linewidth]{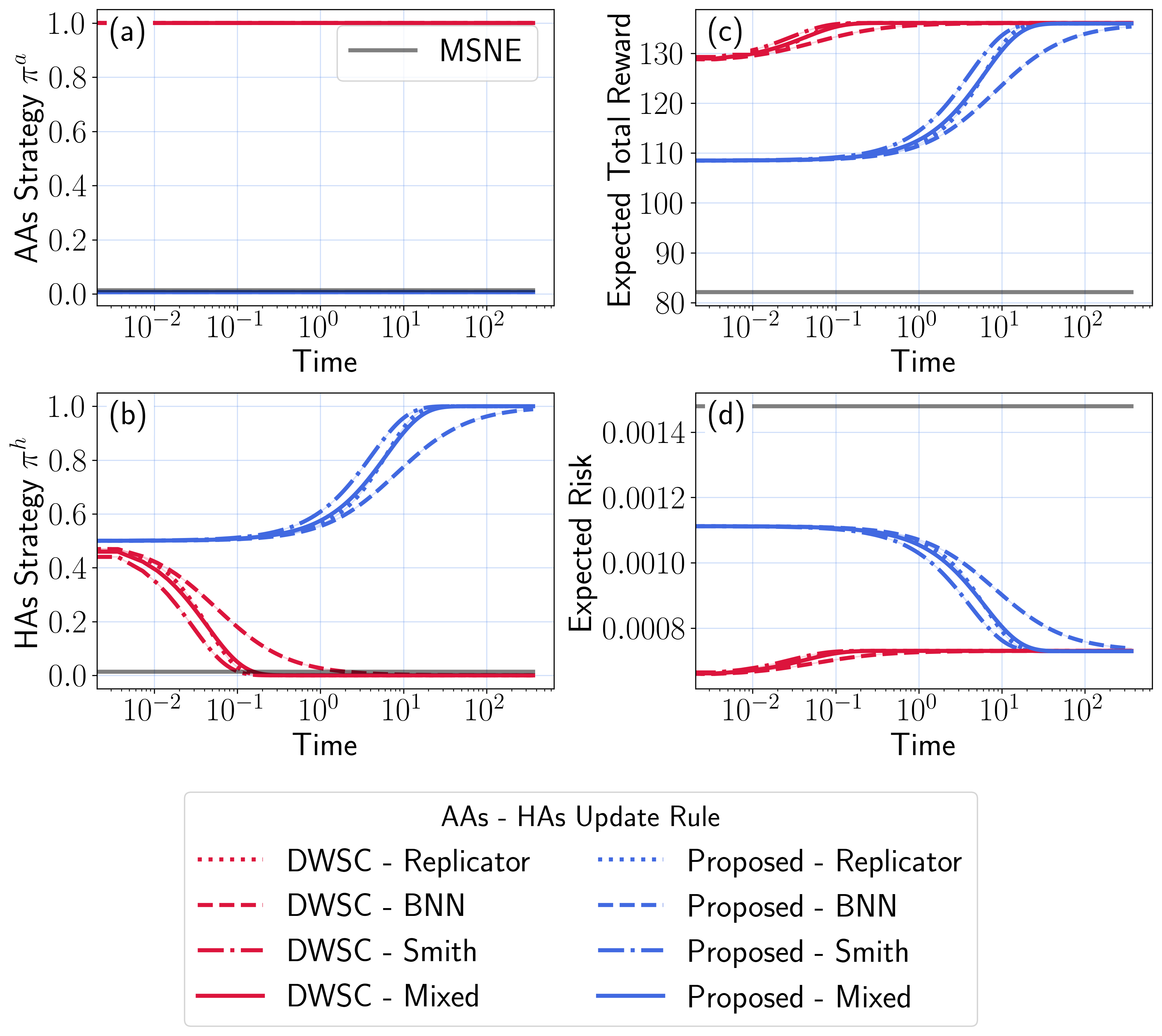}
    \caption{Evolution of states in Type C interaction with different AAs-HAs Update Rules ($\pi^h(0)=0.5$ and $\epsilon=7.29e-4$).}
    \label{fig:res_type_C}
\end{figure}
\begin{figure}[h]
    \centering
    \includegraphics[width=0.6\linewidth]{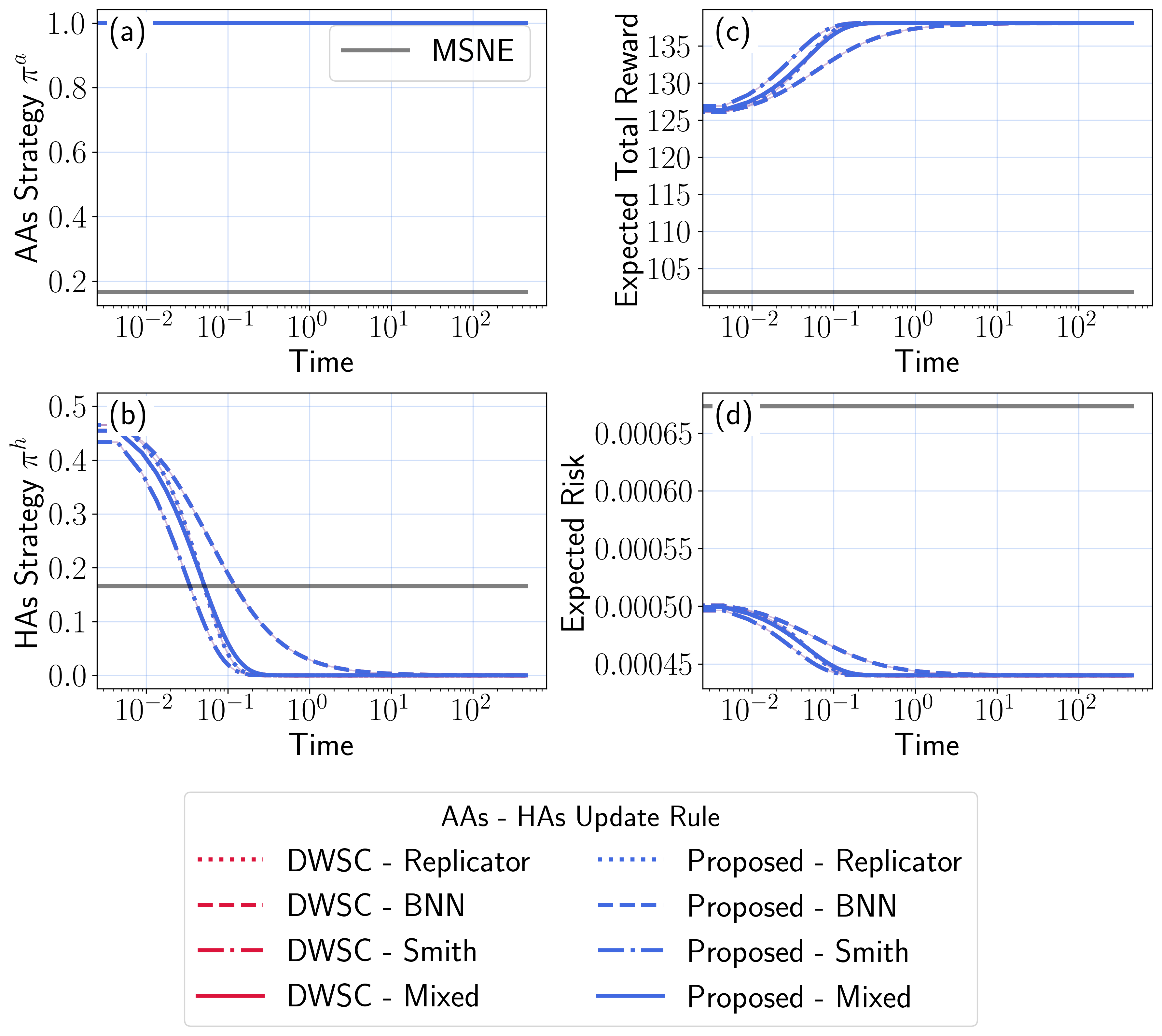}
    \caption{Evolution of states in Type D interaction with different AAs-HAs Update Rules ($\pi^h(0)=0.5$ and $\epsilon=4.4e-4$).}
    \label{fig:res_type_D}
\end{figure}

\begin{figure}[h]
    \centering
    \includegraphics[width=0.6\linewidth]{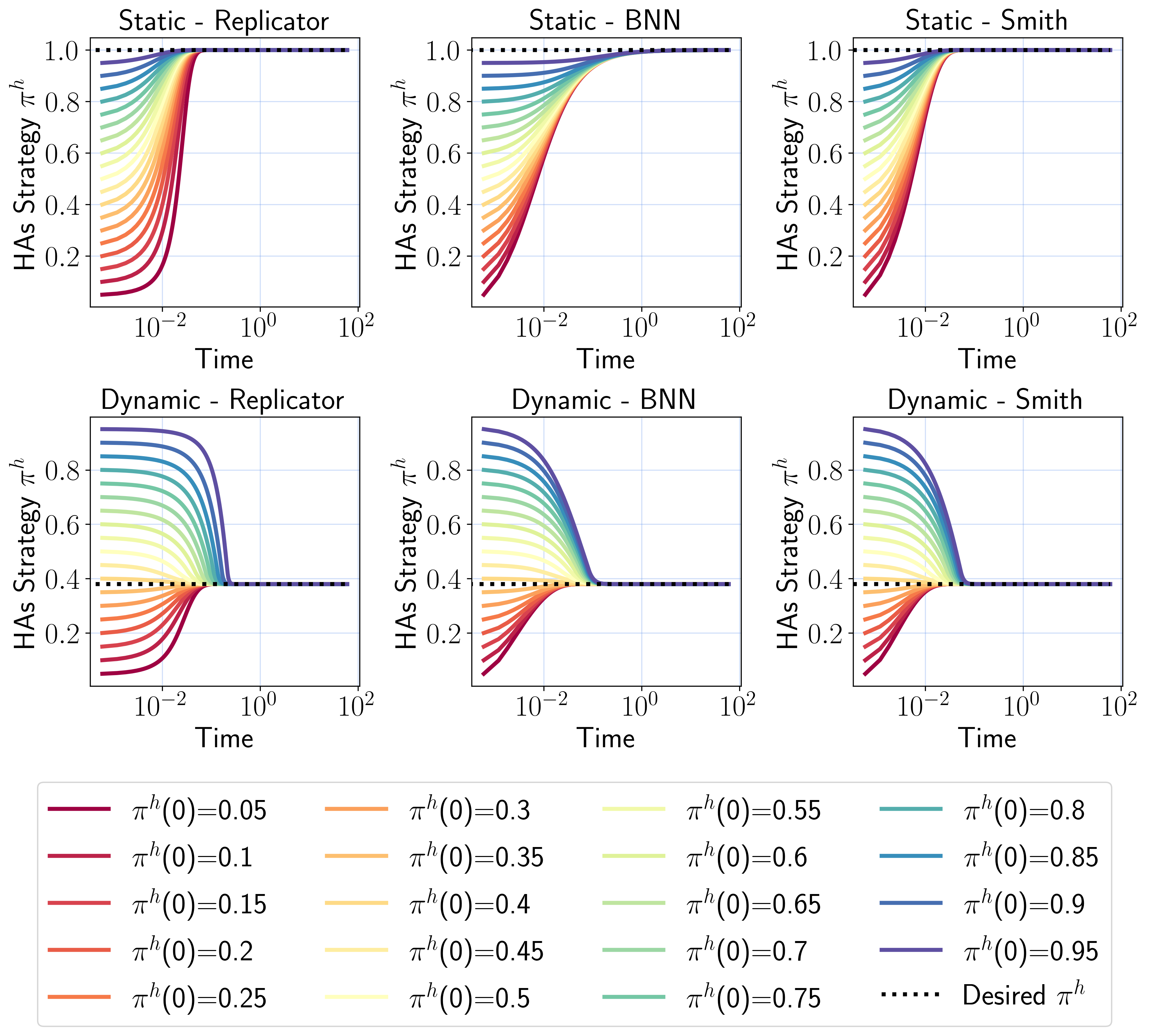}
    \caption{Convergence of HAs' strategy with the proposed methods and uniform HAs' update rules given different initial conditions in Type B game ($\epsilon=1.4e-3$ and $G=1$).}
    \label{fig:res_obj_2a1}
\end{figure}
\begin{figure}[h]
    \centering
    \includegraphics[width=0.6\linewidth]{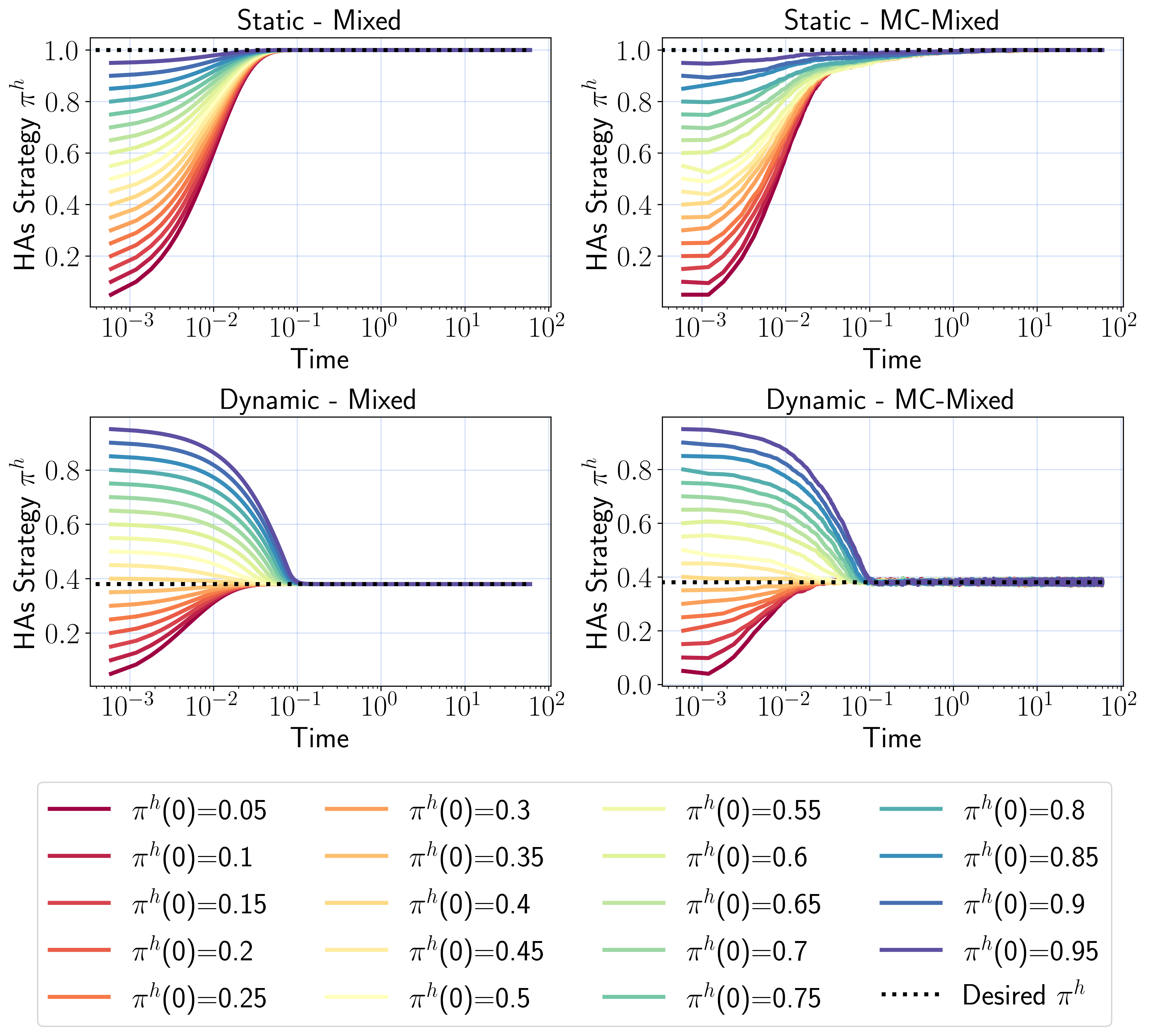}
    \caption{Convergence of HAs' strategy with the proposed methods and mixed HAs' update rules given different initial conditions in Type B game ($\epsilon=1.4e-3$, $w_r=w_b=w_s=\frac{1}{3}$ and $G=1$).}
    \label{fig:res_obj_2a2}
\end{figure}
\begin{figure}[h]
    \centering
    \includegraphics[width=0.6\linewidth]{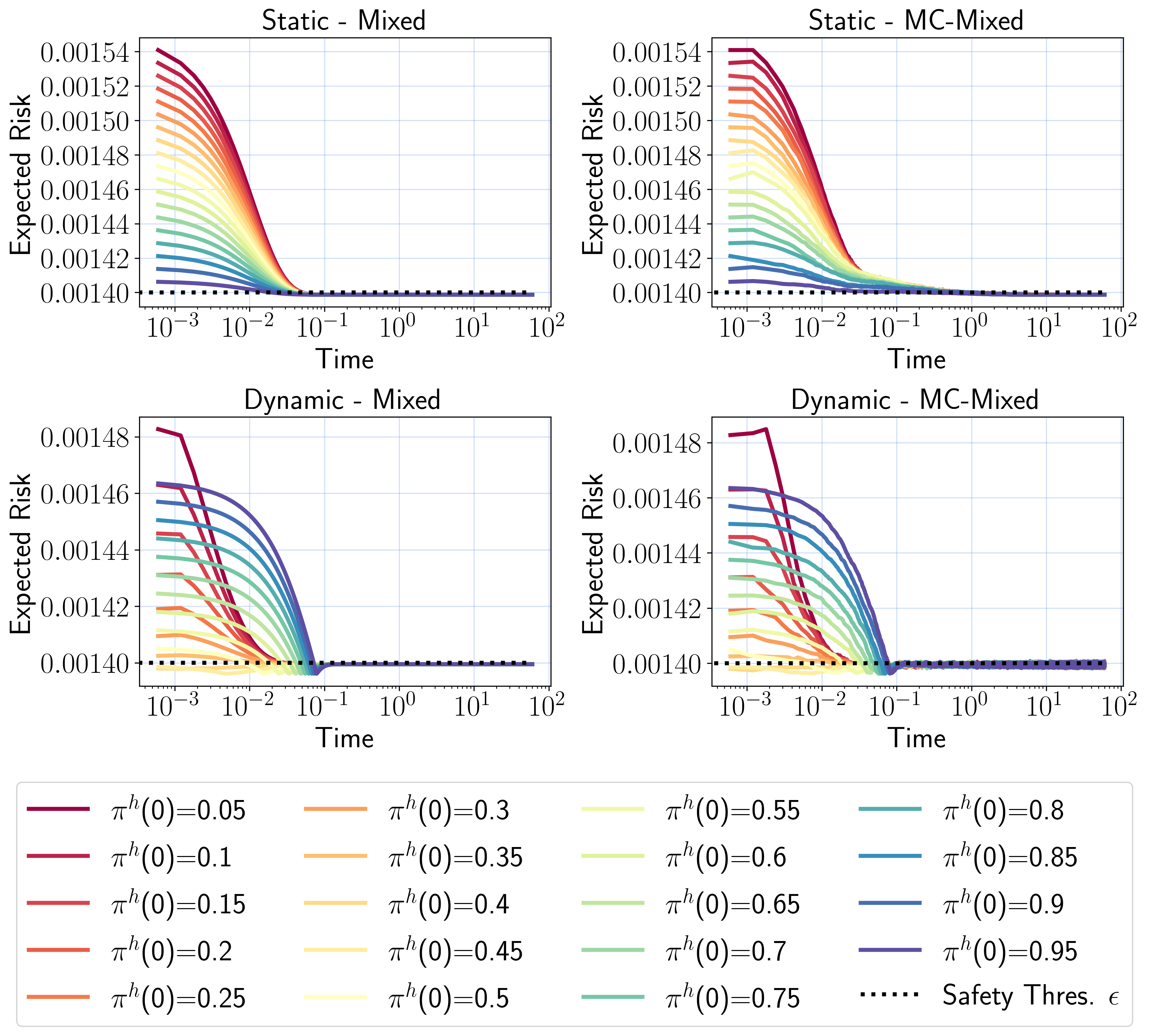}
    \caption{Robustness of the proposed methods controlling risks within a tolerable range against mixed HAs' update rules given different initial conditions in Type B game ($\epsilon=1.4e-3$, $w_r=w_b=w_s=\frac{1}{3}$ and $G=1$).}
    \label{fig:res_obj_2a3}
\end{figure}

Then, we show the vector flow of the strategy of HAs and the admissible strategy $\mathcal{A}$ of Type A interactions with four different dynamics in Figure \ref{fig:admissible_strategy_type_A}. We can see that with $\pi^a<L$, $\dot{\pi^h}>0$ and with $\pi^a>L$, $\dot{\pi^h}<0$, which matches the  \eqref{eq:greater_than_L} and \eqref{eq:less_than_L}.

After that, we present the results of comparison between different AAs’ update rules against different HAs’ update rules in Type B, C, D interactions in \ref{fig:res_type_B} to 
\ref{fig:res_type_D}. 

As shown in Figure \ref{fig:res_type_B} 
for Type B, DWSC achieves the safest strategy while it has the lowest total reward. MSNE has the best reward but worst safety. Here the proposed method is able to achieve the lowest risk but we made a trade-off here so it obtains higher reward than DWSC while keeps the risk below $\epsilon=1.4e-3$.

As shown in Figure \ref{fig:res_type_C}
for Type C, Both DWSC and the proposed method achieve the safest strategy with the highest reward. MSNE has the lowest reward and highest risk at the same time. Theoretically, the proposed method can also made the trade-off to achieve an even safer strategy in Type C interactions. However, due to the parameters used in this setting, the safety advantage is negligible.

As shown in Figure \ref{fig:res_type_D}
for Type D, the proposed method has the same trajectory with DWSC, and both achieve the safest strategy with the highest reward. MSNE has the lowest reward and highest risk at the same time.

To summarize by types, 
\begin{itemize}
    \item Type A: Both the highest rewards and the lowest risk can be achieved at $\pi\in\{(\pi^a,\pi^h)|\pi^a=L,\pi^h=1\}$, which can be achieved by the proposed method only.  
    \item Type B: The highest rewards can be achieved at $\pi\in\{(\pi^a,\pi^h)|\pi^a=L,\pi^h=1\}$, which can be achieved by the proposed method only. The lowest risk can be achived at $\pi\in\{(\pi^a,\pi^h)|\pi^a=0,\pi^h=1 \text{ or } \pi^a=1,\pi^h=0\}$, which can be achieved by the proposed method and DWSC.
    \item Type C: The highest rewards can be achieved at $\pi\in\{(\pi^a,\pi^h)|\pi^a=0,\pi^h=1 \text{ or } \pi^a=1,\pi^h=0\}$, which can be achieved by the proposed method and DWSC. The lowest risk can be achived at $\pi\in\{(\pi^a,\pi^h)|\pi^a=L,\pi^h=1\}$, which can be achieved by the proposed method only.
    \item Type D: Both the highest rewards and the lowest risk can be achieved at $\pi\in\{(\pi^a,\pi^h)|\pi^a=0,\pi^h=1 \text{ or } \pi^a=1,\pi^h=0\}$, which can be achieved by both the proposed method and DWSC.
\end{itemize}

Next, we verified the robustness of the proposed method.
Here we refer to $\pi^{\star}\in \mathcal{F}_d$ and $\pi^{\star}\notin \mathcal{F}_d$ in \ref{eq:proposed_policy} as Dynamic and Static, respectively. To rigorously show the robustness of the proposed method, We present the results of Type B games here. Figure \ref{fig:res_obj_2a1} shows the change of $\pi^h$ in three uniform dynamics of proposed algorithm under different initial conditions in ODE simulations. $\pi^h$ converges to $1$ with Static policy, and the precomputed target $\pi^{h \star}$ with Dynamic policy. 

In addition, we verified that the proposed methods can converge with uncertainties in HA's dynamics. {Figure \ref{fig:res_obj_2a2}} shows the desired convergence behaviors of $\pi^h$ in the Mixed dynamics under different initial conditions in both ODE and Monte Carlo simulations. 

Besides convergence of $\pi^h$, we also show that the risk level can be controlled within a preset threshold in {Figure \ref{fig:res_obj_2a3}}.

\end{document}